\documentclass[11pt,a4paper]{scrartcl}

\usepackage[centertags]{amsmath} 
\usepackage{color} 
\usepackage{amsfonts}
\usepackage{amssymb}
\usepackage{amsthm}
\usepackage{MnSymbol}
\usepackage{rotating}
\usepackage{caption}
\usepackage{booktabs}
\usepackage{enumitem}
\usepackage{hyphenat}
\usepackage{hyperref}
\usepackage{wrapfig}
\usepackage{xcolor}

\usepackage[left=2cm,right=2cm,top=1.6cm,
bottom=1.6cm,includeheadfoot]{geometry}

\newcommand{\N}{\mathbb{N}}
\newcommand{\Z}{\mathbb{Z}}
\newcommand{\Q}{\mathbb{Q}}
\newcommand{\R}{\mathbb{R}}

\newcommand{\Zc}{\mathcal{Z}}
\newcommand{\Xc}{\mathcal{X}}
\newcommand{\Uc}{\mathcal{U}}
\newcommand{\Tc}{\mathcal{T}}
\newcommand{\Pc}{\mathcal{P}}
\newcommand{\Vc}{\mathcal{V}}
\newcommand{\Ec}{\mathcal{E}}
\newcommand{\Gc}{\mathcal{G}}
\newcommand{\Nc}{\mathcal{N}}
\newcommand{\Dc}{\mathcal{D}}
\newcommand{\Rc}{\mathcal{R}}

\newcommand{\ab}{\boldsymbol{a}}
\newcommand{\xb}{\boldsymbol{x}}
\newcommand{\ub}{\boldsymbol{u}}
\newcommand{\bb}{\boldsymbol{b}}
\newcommand{\cb}{\boldsymbol{c}}
\newcommand{\db}{\boldsymbol{d}}

\newcommand{\hb}{\boldsymbol{h}}
\newcommand{\vb}{\boldsymbol{v}}
\newcommand{\yb}{\boldsymbol{y}}

\newcommand{\zb}{\boldsymbol{z}}
\newcommand{\wb}{\boldsymbol{w}}
\newcommand{\rb}{\boldsymbol{r}}
\newcommand{\ssb}{\boldsymbol{s}}
\newcommand{\tb}{\boldsymbol{t}}

\newcommand{\taub}{\boldsymbol{\tau}}
\newcommand{\zetab}{\boldsymbol{\zeta}}
\newcommand{\xib}{\boldsymbol{\xi}}

\newcommand{\Ab}{\boldsymbol{A}}
\newcommand{\Bb}{\boldsymbol{B}}
\newcommand{\Cb}{\boldsymbol{C}}
\newcommand{\Db}{\boldsymbol{D}}
\newcommand{\Kb}{\boldsymbol{K}}
\newcommand{\Qb}{\boldsymbol{Q}}
\newcommand{\Rb}{\boldsymbol{R}}
\newcommand{\Pb}{\boldsymbol{P}}

\newcommand{\Ob}{\boldsymbol{0}}
\newcommand{\Fb}{\boldsymbol{F}}
\newcommand{\Hb}{\boldsymbol{H}}
\newcommand{\Gb}{\boldsymbol{G}}
\newcommand{\Eb}{\boldsymbol{E}}

\newcommand{\Ib}{\boldsymbol{I}}

\newcommand{\Phib}{\boldsymbol{\Phi}}
\newcommand{\Psib}{\boldsymbol{\Psi}}

\newcommand{\Abc}{\boldsymbol{\mathcal{A}}}
\newcommand{\Bbc}{\boldsymbol{\mathcal{B}}}

\newcommand{\Cbc}{\boldsymbol{\mathcal{C}}}
\newcommand{\Dbc}{\boldsymbol{\mathcal{D}}}

\newcommand{\Enc}{\mathtt{Enc}}
\newcommand{\Dec}{\mathtt{Dec}}

\newcommand{\modu}{\mathrm{mod}}

\newcommand{\proj}{\mathrm{proj}}

\colorlet{colorLightBlue}{cyan!12}

\begin{document}

\begin{center}
{\Huge
\textsf{\textbf{Encrypted control for networked systems\,\,}}\\[4mm]
\LARGE 
\textsf{\textbf{An illustrative introduction and current challenges}}
}

\vspace{6mm}
Moritz Schulze Darup\footnote{M. Schulze Darup is with the  Department of Mechanical Engineering, TU Dortmund University, Germany.}$^{,}$\renewcommand{\thefootnote}{$\ast$}\footnotemark[4], \renewcommand{\thefootnote}{\arabic{footnote}}Andreea B. Alexandru\footnote{A.\,B. Alexandru and G.\,J. Pappas are with the Department
of Electrical and Systems Engineering, University of Pennsylvania, USA.}$^{,}$\renewcommand{\thefootnote}{$\ast$}\footnotemark[4], \renewcommand{\thefootnote}{\arabic{footnote}}Daniel E. Quevedo\footnote{D.\,E. Quevedo is with the School of Electrical Engineering \& Robotics, Queensland University of Technology, Australia.}, and George J. Pappas\footnotemark[2]
\vspace{4mm}

\renewcommand{\thefootnote}{$\ast$}
  \footnotetext[4]{These two authors contributed equally to this work and share the first authorship. Correspondence to:
  
\noindent \texttt{moritz.schulzedarup@tu-dortmund.de} or \texttt{aandreea@seas.upenn.edu}.
        }
\end{center}


\begin{wrapfigure}{r}{0.55\textwidth}
\vspace{-5mm}
\fcolorbox{black}{colorLightBlue}{
  \begin{minipage}{0.52\textwidth}
\textsf{\textbf{Summary.}} Control systems are rapidly evolving to utilize modern computation and communication tools, such as cloud computing and geographically distributed networks, in order to improve performance, coverage and scalability. However, control loops that outsource the computation over privacy-sensitive data to third-party platforms via public networks are already the subject of cyberattacks involving eavesdropping and data manipulation. 
Encrypted control addresses this security gap and provides confidentiality of the processed data in the entire control loop, by encrypting the data at each level of transmission (over the network) and of computation (on corrupted computing platforms). 
This paper presents a tutorial-style introduction to this young but emerging field in the framework of secure control for networked dynamical systems with encrypted data. We focus on the steps of deriving the encrypted formulations of some specific control algorithms from the standard formulations and discuss the challenges arising in this process, ranging from privacy-aware conceptualizations to changes in the computation flows and quantization issues. In conclusion, we provide a list of open problems and new directions to explore in order to consolidate the area of encrypted control. 
\end{minipage}
}\vspace{-5mm}
\end{wrapfigure}

Cloud computing and distributed computing are becoming ubiquitous in many modern control systems such as smart grids, building automation, robot swarms or intelligent transportation systems. Compared to ``isolated'' control systems, the advantages of cloud-based and distributed control systems are, in particular, resource pooling and outsourcing, rapid scalability, and high performance. However, these capabilities do not come without risks. In fact, the involved communication and processing of sensitive data via public networks and on third-party platforms promote, among other cyberthreats, eavesdropping and manipulation of data. 
That these threats are relevant to real-world applications is apparent from an increasing number of cyberattacks explicitly addressing industrial control systems~\cite{hackmageddon2019statistics}. Prominent examples are the malwares Stuxnet, Duqu, Industroyer, or Triton (see, e.g.,~\cite{Chen2011_Stuxnet}) as well as inference attacks arising from smart meters used as surveillance devices (see, e.g. ~\cite{Molina2010private,Greveler2012multimedia}). Clearly, cyberattacks on control systems can be highly critical. In particular, unlike attacks on classical IT systems, attacks on control systems may influence physical processes through digital manipulations~\cite{Teixeira2015_CSM}.
Moreover, networked control systems are the backbone of critical infrastructure such as electric power, transportation, and water distribution networks, with further applications illustrated in the sidebar ``\hyperlink{side:applications}{Prospective uses of encrypted control in industry}''.  
Hence, future control schemes should  counteract privacy and security threats and ensure confidentiality, integrity, and availability (see \cite{Bishop2002} or \cite[Fig.~S1]{Teixeira2015_CSM} for details on these traditional security goals) of the involved process data.

Secure control for networked systems has been intensively studied in the literature during the last decade. Comprehensive surveys can be found in~\cite{Cardenas2008_DCSW,Pasqualetti2013_TAC,Knowles2015_CIP,Chong2019_ECC}
and in the special issue of the IEEE Control Systems Magazin on ``Cyberphysical Security'' from 2015  (especially \cite{Teixeira2015_CSM}).
Two observations are particularly important for the scope of this paper. First, analogously to cyberattacks on IT systems, there exists a variety of different attacks and tailored defense mechanisms. For instance, stealth, false-data injection, replay, covert, and denial-of-service (DoS) attacks can be distinguished~\cite{Pasqualetti2013_TAC}. 
Second, interdisciplinary solutions are required to secure control systems. In fact, standard information-theoretic or cryptographic tools on their own are often not sufficient (see, e.g.,~\cite[Sect.~3]{Cardenas2008_DCSW} or~\cite{vanDijk2010_USENIX}). 
Most existing works focus on the integrity and availability of networked control schemes using various defense mechanisms. For example, control-related concepts such as detectability and identifiability of deception attacks are investigated in~\cite{Pasqualetti2013_TAC}. Moreover, game-theoretic approaches to deal with DoS attacks are, e.g., considered in~\cite{Gupta2010_CDC,Li2015_TAC}.

\vspace{3mm}
\begin{wrapfigure}{l}{0.6\textwidth}
\vspace{-4mm}
\fcolorbox{black}{colorLightBlue}{
  \begin{minipage}{0.56\textwidth}
\textsf{\textbf{\hypertarget{side:applications}{Prospective uses of encrypted control in \nohyphens{industry}}.}} Recent years saw the emergence of ``Control-as-a-Service"~\cite{Givehchi2014control} with companies offering personalized optimized control algorithms for a better power consumption and economic efficiency. Encrypted control, which deals with privately computing control decisions while processing encrypted data, is an appealing solution to the types of attacks mentioned in the Introduction. Control-as-a-Service is mainly used in building automation and smart grids, but took off also in automation systems for manufacturing and chemical industries, food service, transportation systems, water supply, and sewage waste maintenance. The inertia of such systems makes encrypted control (which has an overhead compared to standard control due to encryption) particularly amenable. Furthermore, given the criticality of some of these industrial systems, control schemes that simultaneously take into account security, safety, and constraints, such as encrypted model predictive control, are recommended. Apart from these examples, lighter-weight distributed encrypted control could be adequate in exploration and surveillance swarms of robots deployed in hazardous environments, where the capture of one robot that operates an encrypted control scheme would not jeopardize the other robots in the swarm.
\end{minipage}
}\vspace{-4mm}
\end{wrapfigure}

\begin{figure}[htp]
\centering
\includegraphics[width=0.98\textwidth]{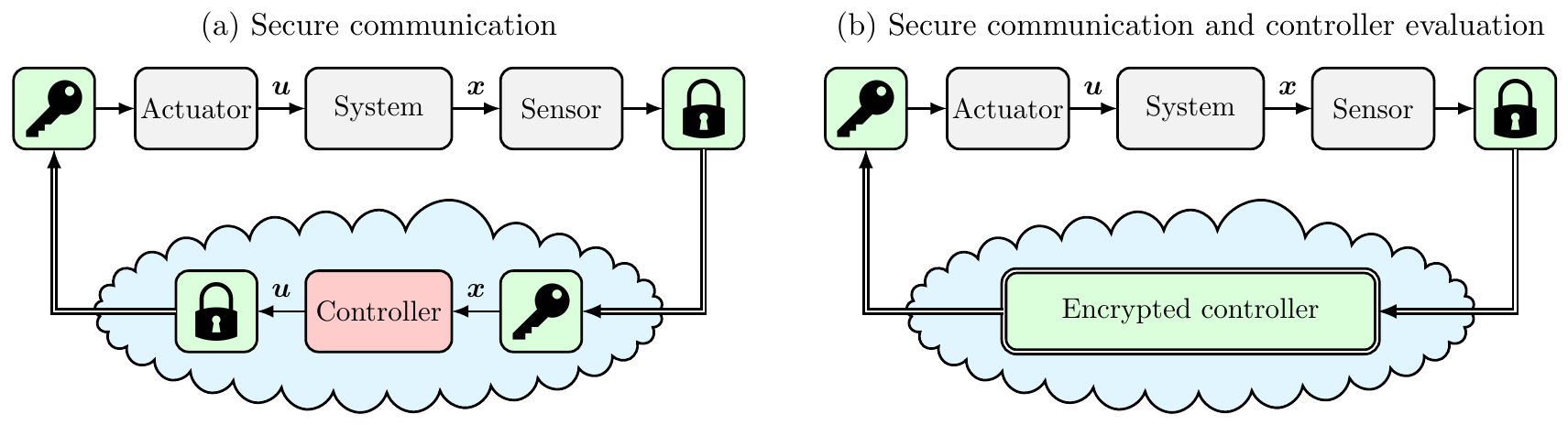} 
\caption{Illustration of (a) a cloud-based control scheme with encrypted communications but insecure
controller evaluation and (b)  a cloud-based control scheme with encrypted communications and encrypted controller evaluation. Double-arrows and double-framing highlight encrypted data transmission and encrypted data processing, respectively.}
\label{fig:encryptedControl}
\end{figure}

The emerging field of encrypted control primarily aims for confidentiality of sensitive system states, control actions, controller parameters, or model data in the entire control loop. 
More generally, an encrypted controller can be defined as a networked control scheme that simultaneously ensures control performance and privacy of the client system(s) through specialized cryptographic tools. 
In the framework of networked control, attacks compromising confidentiality such as eavesdropping might seem less critical since they do not immediately cause physical misbehavior. However, ``passive'' spying often precedes ``active'' attacks compromising data integrity and availability (see, e.g.,~\cite[Sect.~III.A]{Chong2019_ECC}). 
Abstractly speaking, encrypted control is realized by modifying conventional control schemes such that they are capable of computing encrypted inputs based on encrypted states (or encrypted controller parameters) without intermediate decryptions by the controller. The basic concept is illustrated in Figure~\ref{fig:encryptedControl}.(b) for a cloud-based controller. 
In this context, it is important to note that encrypted control goes beyond secure communication channels as in Figure~\ref{fig:encryptedControl}.(a) that could be realized using classical encryption, such as AES.
In fact, encrypted control additionally provides security against curious cloud providers or neighboring agents that, during controller evaluations, would have access to unsecured data for solely secured communications. 
In this context, the consideration of so-called honest-but-curious platforms is the key difference to existing secure control schemes focusing on confidentiality (such as, e.g.,~\cite{Tsiamis2017_CDC,Leong2017_IFAC}) and it is part of the attack model underlying most encrypted control schemes as specified in the sidebar ``\hyperlink{side:securityAttacks}{Security against what? Security goals and attack models}''.
Meeting these privacy demands under real-time restrictions is non-trivial and requires a co-design of controllers and suitable cryptosystems. In fact, identifying controller formulations that can be efficiently combined with capable cryptosystems can be seen as the central task for encrypted control and will be a recurring theme throughout this paper. 

\begin{figure}[h!]
\vspace{-4mm}
\fcolorbox{black}{colorLightBlue}{
  \begin{minipage}{0.96\textwidth}
\textsf{\textbf{\hypertarget{side:securityAttacks}{Security against what? Security goals and attack models}.}} In modern cryptography, the concept of ``security'' is always related to a security goal and certain attacks. Roughly speaking, an encryption scheme
is considered secure against disclosure attacks if an attacker gets no new information about a plaintext (i.e., the unencrypted data) from a ciphertext (i.e., the encrypted data) -- regardless of its previous knowledge~\cite[p.~19]{Katz2014}.
The knowledge and capabilities of an attacker are the basis for the attack model. For example, one can distinguish between ciphertext-only attacks, where the attacker just observes ciphertexts and attempts to obtain information about the underlying plaintexts, and known-plaintext attacks, where the attacker has access to some plaintext-ciphertext tuples~\cite[p.~20]{Katz2014}. Further variants of the latter attack are chosen-plaintext attacks, where the attacker gets access to ciphertexts for plaintexts of its choice, and chosen-ciphertext attacks, where the attacker gets access to plaintexts for ciphertexts of its choice.
We next specify some popular security goals. In the framework of cryptography, the highest security level is perfect secrecy~\cite{Shannon1949}. 
Roughly speaking, a cryptosystem is called perfectly secret if the probability distribution of the ciphertext does not depend on the plaintext (cf. \cite[p.~29]{Katz2014}). In other words, the probability of one message to be the one encrypted is the same for all possible messages in the space.
The easiest way to specify this definition is to study a simple scheme that is not perfectly secret. Consider, e.g., the substitution cipher in the table below and assume it is used to encrypt an English text. 
In this example, the probability distribution of the plaintext reflects the well-known average letter frequencies for English text (see, e.g.,~\cite[Fig.~1.3]{Katz2014}). Now, due to the fixed substitutions, the frequencies of the ciphertexts depend on the plaintext. Hence, the scheme (that is a generalization of Caesar's cipher) is not perfectly secret. Perfect secrecy against ciphertext-only attacks is, however, provided by the so-called one-time pad that is briefly discussed in Section~\ref{sec:Basics}. 

\begin{center}
\small
\vspace{-1.5mm}
\begin{tabular}{lcccccccccccccccccccccccccc}
\toprule
Plain & a\!\! & b\!\! & c\!\! & d\!\! & e\!\! & f\!\! & g\!\! & h\!\! & i\!\! & j\!\! & k\!\! & l\!\! & m\!\! & n\!\! & o\!\! & p\!\! & q\!\! &r\!\! & s\!\! & t\!\! & u\!\! & v\!\! & w\!\! & x\!\! & y\!\! & z\!\! \\
Cipher\!\!\!\! & E\!\! & N\!\! & C\!\! & R\!\! & Y\!\! & P\!\! & T\!\! & G\!\! & O\!\! & L\!\! & D\!\! & F\!\! & I\!\! & S\!\! & H\!\! & M\!\! & U\!\! & Z\!\! & A\!\! & K\!\! & B\!\! & J\!\! & Q\!\! & V\!\! & W\!\! & X\!\!  \\
\bottomrule
\end{tabular}
\vspace{-1.5mm}
\end{center}

As apparent from its definition, perfect secrecy is an information-theoretic concept. Hence, it does not depend on the difficulty of some computational problems. 
In contrast, the security of most homomorphic cryptosystems is called semantic security, which roughly means that the ciphertexts are computationally indistinguishable from each other, and builds on the hardness of certain computational problems.
For example, the semantic security of ElGamal~\cite{ElGamal1985} requires that the discrete logarithm is difficult to compute (in the underlying group). Analogously, the semantic security of Paillier~\cite{Paillier1999} builds on the decisional composite residuosity assumption, which is related to the hardness of factorization~\cite[p.~495]{Katz2014}. For such schemes, it is natural to include computational properties also in the security specification. 
In contrast to perfect secrecy, semantic security specifies that some information about the plaintext can be extracted from the ciphertext.
However, the revealed information should either be negligible or extracting the information should be computationally infeasible (see, e.g.,~\cite[Def.~3.12]{Katz2014} for a more specific definition). Under the assumption that the underlying computational problems are hard, ElGamal and Paillier are both semantically secure against chosen-plaintext attacks. Interestingly, due to their homomorphisms, they are inherently not secure against chosen-ciphertext attacks. 
In encrypted control, we mainly consider attacks that focus on disclosing secret information, e.g., through eavesdropping. We further assume that attacks can occur during communication but also in the cloud(s) or at neighboring agents. It is, however, assumed that 
these parties follow prescribed protocols. In other words, the external parties honestly evaluate outsourced algorithms but they are curious about the processed data. This attack model is known as honest-but-curious and it is standard in cryptography (see, e.g.,~\cite[Ch.~2]{Evans2018pragmatic}). In addition, honest-but-curious cloud providers are realistic, since it is profitable for them to learn and use the private information of users, but it is extremely risky for their business model to deviate from the a priori established protocols and be caught. 
\end{minipage}}\vspace{-5mm}
\end{figure}

\newpage

Encrypted control was first realized in~\cite{Kogiso2015} for linear state feedback. Since then, a few dozen papers have presented more efficient implementations and encrypted realizations of more complex control schemes (see the sidebar~``\hyperlink{side:literatureSurvey}{A brief survey on encrypted control}'' for details).
The different approaches can be classified based on controller type, system architecture, and applied cryptographic techniques. So far, linear, model predictive, and distributed controllers have been implemented in an encrypted fashion. 
Regarding the architecture, we can distinguish networked control systems including one cloud, multiple (non-collaborating) clouds, or multiple agents (cf.~Figs.~\ref{fig:encryptedControl}, \ref{fig:twoCloudsSharing}, and \ref{fig:encryptedCooperation}). 
Concerning the cryptographic tools, solutions involving homomorphic encryption, secret sharing, and hybrid methods that ensure secure multi-party computation have been proposed.
Since the mentioned cryptosystems may be new to many readers, we briefly summarize the underlying concepts before giving formal descriptions further below. Homomorphic encryption (see, e.g.,~\cite{ElGamal1985,Paillier1999,Gentry2010,Brakerski2014}) refers to a special family of cryptosystems that allow elementary mathematical operations (such as additions or multiplications) to be carried out on encrypted data. Secret sharing~\cite{Shamir1979} divides secret data into multiple shares, such that individual shareholders learn nothing about the secret but the secret can be reconstructed by combining a certain number of shares. Secure multi-party computation (see, e.g.,~\cite{Yao1982,Pinkas2009}) provides protocols to perform computations on secret data, e.g., based on secret shares and oblivious transfer.

\vspace{2mm}
\begin{figure}[h]
\fcolorbox{black}{colorLightBlue}{
  \begin{minipage}{0.96\textwidth}
\textsf{\textbf{\hypertarget{side:literatureSurvey}{A brief survey on encrypted control}.}} Encrypted linear controllers, including linear-quadratic (Gaussian) regulators, implemented in a single cloud using partially homomorphic encryption HE have been proposed in~\cite{Fujita2015,Kogiso2015,Farokhi2016,Farokhi2017}, \cite{Lin2018secure,Cheon2018need,Teranishi2019stability,Murguia2020}, leveled HE in~\cite{Alexandru2019_CPS} and fully HE in~\cite{Kim2016}. 
An encrypted linear controller utilizing two clouds, secret sharing, and secure two-party computation has recently been discussed in~\cite{SchulzeDarup2019_CDC_2Clouds}. A homomorphically encrypted and authenticated linear control has been proposed in~\cite{Cheon2018toward} using labeled HE. Encrypted model predictive control (MPC) using a single cloud and additively HE is addressed in~\cite{SchulzeDarup2018_LCSS,SchulzeDarup2018_NMPC,SchulzeDarup2020}.
Encrypted MPC using two clouds and a combination of HE and secure multi-party computation are presented in~\cite{Alexandru2018_CDC,Alexandru2020}. 
The private Kalman filtering problem is considered in~\cite{Gonzalez14}, where a solution involving secure multi-party computation and HE is proposed. An encrypted controller for discrete event systems is investigated in~\cite{Fritz2019controller}. 
Finally, encrypted distributed control for multi-agent systems is considered in~\cite{Kishida2018_CDC,SchulzeDarup2019_LCSS,Alexandru2019_CDC,Ruan2019_TAC}. 
An important side note is that of encrypted optimization, that can be further used to design encrypted controllers, for example MPC. Works that have proposed schemes for solving optimization problems on encrypted data include~\cite{Shoukry2016,Alexandru2017_Allerton,Lu2018_Auto,Tjell2019_CDC}. 
The approaches in~\cite{Kishida2018_CDC,SchulzeDarup2019_LCSS,Ruan2019_TAC} mainly build on additively HE whereas secure multi-party computation dominates in~\cite{Alexandru2019_CDC,Tjell2019_CDC}. 
\end{minipage}
}
\end{figure}

This article provides an illustrative introduction to the young but emerging field of encrypted control for non-cryptographers. We discuss and unify the different types of existing encrypted controllers, their architectures, and the underlying cryptographic tools. 
In particular, we summarize the linear controllers in~\cite{Kogiso2015,Farokhi2016,SchulzeDarup2019_CDC_2Clouds} 
that build on multiplicatively homomorphic encryption, additively homomorphic encryption, and secure two-party computation, respectively.
We further investigate the encrypted model predictive controlles in~\cite{SchulzeDarup2018_LCSS,SchulzeDarup2018_NMPC,Alexandru2018_CDC} and the encrypted cooperative control schemes in~\cite{SchulzeDarup2019_LCSS} and~\cite{Alexandru2019_CDC}. The selected works allow us to illustrate two core ingredients of encrypted controllers. First, we present fundamental cryptographic tools that enable encrypted computations. Second, we discuss reformulations of conventional control schemes that support the application of those tools.
In addition, we indicate current challenges in encrypted control arising from the presented works and thus provide stimuli for future research. 
The outline of this article is as follows. We first provide basics on homomorphic encryption, secret sharing, and multi-party computation. Afterwards, we discuss milestones of encrypted control based on realizations of linear, model predictive, and distributed controllers. Finally, we point out current challenges and future trends in encrypted control.

\section{Basics on homomorphic encryption, secret sharing,\\ \quad and multi-party computation}
\label{sec:Basics}

In what follows, we call \textit{plaintext} a piece of data that is unencrypted. Plaintexts can refer to both decrypted data as well as data that is prepared for encryption, e.g., a quantized number. In a similar manner, we call \textit{ciphertext} an encrypted piece of data, that, without secret information such as the private key, cannot be used to retrieve the plaintext that was encrypted. 

Essentially, homomorphic encryption (HE) refers to a special family of cryptosystems that enables mathematical operations to be carried out on encrypted data (see, e.g,~\cite{ElGamal1985,Paillier1999,Gentry2010,Brakerski2014,Cheon2017_CKKS}). 
Let $z_1$ and $z_2$ be two arbitrary numbers in the message space of the cryptosystem and denote by ``$\Enc$'' and ``$\Dec$'' the encryption and decryption procedure, respectively.
More precisely, we call a cryptosystem multiplicatively homomorphic if there exists an operation ``$\otimes$'' that allows the evaluation of encrypted multiplications, i.e., the relation
\begin{equation}
\label{eq:multiplicativelyHomomorphic}
z_1 \, z_2 = \Dec \left( \Enc(z_1) \otimes \Enc(z_2) \right)
\end{equation}
holds.
Analogously, cryptosystems are called additively homomorphic if an operation ``$\oplus$'' exists such that encrypted additions can be realized:
\begin{equation}
\label{eq:additivelyHomomorphic}
z_1 + z_2 = \Dec \left( \Enc(z_1) \oplus \Enc(z_2) \right).
\end{equation}
A popular multiplicatively HE scheme is the ElGamal cryptosystem~\cite{ElGamal1985}. Additively HE is often implemented using Paillier (see~\cite{Paillier1999} or the sidebar ``\hyperlink{side:Paillier}{Paillier encryption and its homomorphisms}'').
Encryption schemes that are both additively and multiplicatively homomorphic are called fully homomorphic. In principle, fully HE schemes can be used to encrypt arbitrary functions~\cite{Gentry2010}. However, fully HE is computationally highly demanding and currently not a competitive option for encrypted control on power or memory constrained devices.
Nevertheless, so-called somewhat or leveled fully HE schemes (see, e.g.,~\cite{Brakerski2014,Cheon2017_CKKS}),
that support a limited number of encrypted multiplications and (a limited or unlimited number of) encrypted additions, have already been used for encrypted control and may become more relevant in the future.

If multiple cloud servers or agents are involved in the control law evaluations, secret sharing combined with other multi-party computation techniques are an option to realize encrypted control. Roughly speaking, secret sharing allows to divide and recombine secret data in such a way that the individual shares reveal no information about the secret~\cite{Shamir1979}.  
More formally, a secret sharing scheme that splits secret data into $S\geq 2$ shares in such a way that at least $L \in \{2,\dots,S\}$ shares are required to reconstruct the secret is called an $(L,S)$-threshold scheme.

\begin{wrapfigure}{r}{0.65\textwidth}
\vspace{-1mm}
\fcolorbox{black}{colorLightBlue}{
  \begin{minipage}{0.61\textwidth}
\textsf{\textbf{\hypertarget{side:Paillier}{Paillier encryption and its homomorphisms}.}} So far, the majority of encrypted controllers makes use of the Paillier cryptosystem~\cite{Paillier1999}. 
The standard realization of the scheme considers the additive group of integers modulo $P$, denoted as $\Z_P$, as the message space, where $P$ is the product of two large primes $p$ and $q$. For the purposes of this paper, it is convenient to think of $\Z_P$ as the set $\{0,\dots,P-1\} \subset \N$. 
During the key generation procedure, two large primes $p, q \in [2^{\ell-1},2^\ell]$ of the same ``length'' $\ell \in \N$ are selected and the public key is specified by $P=pq$.
The secret key, that is only known to the entity that needs to decrypt the data, is computed as $S=(p-1)(q-1)$.
Now, the encryption of a number $z$ from $\Z_P$ is realized by
\begin{equation}
\label{eq:encPaillier}
\Enc(z,r):=(P+1)^z r^P \,\modu \,\, P^2,
\end{equation}
where $r$ is randomly picked from $\Z_{P^2}^\ast$, the multiplicative group of integers modulo $P^2$. The resulting ciphertext $c$ is likewise an element of $\Z_{P^2}^\ast$ and the decryption is carried out by computing
\begin{equation}
\label{eq:decPaillier}
\Dec(c):=\frac{(c^S \, \modu \,P^2) - 1}{P}\, \, \mu  \,\,\modu \,\,  P
\end{equation} 
using the secret key $S$, where $\mu$ is the multplicative inverse of $S$ modulo $P$.
It is easy to prove correctness of the scheme, i.e.,  $z=\Dec(\Enc(z,r))$ for every $z \in \Z_P$ and every $r \in \Z_{P^2}^\ast$.
Semantic security of the Pallier scheme builds on the assumption that the decisional composite residuosity problem (see, e.g.,~\cite[p.~495]{Katz2014} for details) is hard to solve, which is currently the case for large $P$ (thousands of bits).
Regarding homomorphisms, we can infer from \eqref{eq:encPaillier} and \eqref{eq:decPaillier} that the Paillier cryptosystem is additively homomorphic. In fact, we have 
\begin{equation}
\label{eq:addPaillier}
z_1+z_2 = \Dec\left(\Enc(z_1,r_1) \,\Enc(z_2, z_2)\,\,\modu \,\,P^2 \right)
\end{equation}
for all $z_1,z_2 \in \Z_P$  and all $r_1,r_2 \in \Z_{P^2}^\ast$. At this point, we recall that the messages are integers. Hence, multiplications $z_1 z_2$ can be implemented as sums of the form
\begin{equation}
\label{eq:integerProductAsSum}
z_1 z_2 = \underbrace{z_1+\dots+z_1}_{z_2\text{-times}}.
\end{equation}
Applying the homomorphism~\eqref{eq:addPaillier} repeatedly to
the right-hand side of~\eqref{eq:integerProductAsSum} yields the relation
\begin{equation}
\label{eq:mulCPaillier}
z_1 z_2 = \Dec\left(\Enc(z_1,r)^{z_2} \,\,\modu \,\, N^2\right)
\end{equation}
that holds for all $z_1,z_2 \in \Z_P$ and all $r \in \Z_{P^2}^\ast$.
Relation~\eqref{eq:mulCPaillier} allows to implement partially encrypted multiplications with one encrypted factor (here $z_1$).  
This feature is an advantage of additively homomorphic cryptosystems  over multiplicatively homomorphic schemes.
\end{minipage}
}\vspace{-1mm}
\end{wrapfigure}

In the simplest case, $S=2$ shareholders are involved, which implies $L=2$, i.e., 
the secret data is split into two shares and both shares are required to reconstruct the secret. As an example, let us assume the secret is an integer $z \in \{0,\dots,\varphi-1\}$ for some user-defined $\varphi \in \N$ with $\varphi\geq 2$.
Then, a $(2,2)$-threshold scheme can be realized as follows.  
An integer $r \in \{0,\dots,\varphi-1\}$ is chosen uniformly at random and the~\nohyphens{ciphertext}
\begin{equation}
\label{eq:encOneTimePad}
c :=  z + r  \,\, \modu \,\, \varphi 
\end{equation}
is computed, where $\modu$ refers to the standard modulo operation. The two numbers $c$ and $r$ can be considered as two shares of the secret $z$. In this context, it is easy to see that neither $c$ nor $r$ reveal any information on $z$. In fact, $r$ is a random number and $c$ as in~\eqref{eq:encOneTimePad} corresponds to the ciphertext of a so-called one-time pad. 
Since one-time pads are perfectly secure (see \cite{Shannon1949} and the sidebar ``\hyperlink{side:securityAttacks}{Security against what? Security goals and attack models}''), $c$ reveals no information on~$z$ apart from the fact that $z \in \{0,\dots,\varphi-1\}$. Nevertheless, combining the shares $c$ and $r$ allows one to reconstruct $z$ according to
$z = c - r \,\, \modu \,\, \varphi$.

Secure multi-party computation consists of a set of techniques that involve the secure processing of data between multiple mutually distrusting parties~\cite{Evans2018pragmatic}. 
A secure multi-party computation scheme can be realized by splitting the private data among multiple parties using secret sharing, garbled circuits~\cite{Bellare2012}, or (threshold) HE~\cite{Cramer2001multiparty} and then jointly computing a mutually agreed upon function on their pieces of information using hybrid versions between the aforementioned tools.
An example of a protocol essential to secure multi-party computation is (1-out-of-2) oblivious transfer. To illustrate this protocol, consider a party that holds two numbers $z_0,z_1 \in \{0,\dots,\varphi-1\}$ and a second party that has a bit $\tau \in\{0,1\}$. 
The problem underlying oblivious transfer is to enable the second party to obtain $z_\tau$ without finding out the other number (i.e., $z_{1-\tau}$) and without revealing $\tau$ to the first party. There exist various implementations of oblivious transfer (see, e.g., \cite[Ch.~3.7]{Evans2018pragmatic}). We next present a simple realization based on additively HE that we will later use to implement an encrypted model predictive controller.
To prepare the oblivious transfer, the second party sets up a Paillier cryptosystem by generating a public key used for encryption and homomorphic operations and a secret key required for decryption (see the sidebar ``\hyperlink{side:Paillier}{Paillier encryption and its homomorphisms}'' for details). The first party then generates two random numbers $r_0$ and $r_1$, computes $z_0 + r_0  \,\, \modu \,\, \varphi$ and $z_1 + r_1 \,\, \modu \,\, \varphi $ according  to \eqref{eq:encOneTimePad}, and sends the results to the second party. Based on its bit~$\tau$, the second party selects, encrypts and sends back $\Enc(z_\tau + r_\tau  \,\, \modu \,\, \varphi)$ and $\Enc(\tau)$. Next, the first~party~computes
\begin{equation}
\label{eq:obliviousTransfer}
c:=\Enc(z_\tau+r_\tau \,\, \modu \,\, \varphi) \oplus \left((\Enc(\tau) \oplus \Enc(-1) ) \,\odot\, r_0\right)  \oplus \left( \Enc(\tau)  \,\odot\, (-r_1)\right)
\end{equation}
by exploiting~\eqref{eq:additivelyHomomorphic} and another (related) homomorphism of the Paillier scheme that we denote with~``$\odot$''.
In fact, as pointed out in the sidebar ``\hyperlink{side:Paillier}{Paillier encryption and its homomorphisms}'', Paillier not only offers~\eqref{eq:additivelyHomomorphic} but also the relation~\eqref{eq:mulCPaillier} that enables multiplication between one encrypted factor and one plaintext integer (i.e., $\Enc(z_1) \odot z_2$). 
Afterwards, the first party returns the resulting ciphertext $c$ to the second party, which finally obtains $z_\tau$ from evaluating $\Dec(c) \,\, \modu \,\, \varphi$ using its secret key. Obviously, $z_\tau$ is correctly received since~\eqref{eq:obliviousTransfer} actually implements
$$
z_\tau+r_\tau \,\, \modu \,\, \varphi +r_0(\tau-1) -  r_1 \tau  = z_\tau+r_\tau \,\, \modu \,\, \varphi - r_\tau.
$$
The privacy of the oblivious transfer follows from the security of the additively homomorphic cryptosystem and the statistical security of the secret sharing scheme.

\vspace{-1mm}
\begin{figure}[h]
\vspace{-1mm}
\fcolorbox{black}{colorLightBlue}{
  \begin{minipage}{0.96\textwidth}
\textsf{\textbf{\hypertarget{side:SoftwareHardware}{Software for encrypted computations and hardware acceleration}.}} One of the goals of encrypted computation is to provide a wide-range usage, which requires optimized code and versatility of functions. To this end, there are several open-source software packages available. Regarding partially homomorphic encryption (HE), a multiplicity of implementations exists due to their relatively simple functionality. 
We exemplarily refer to \texttt{python-paillier}~\cite{PythonPaillier} that implements the Paillier scheme, with the security caveat that the default encoding is a floating-point encoding with public exponent. For fully HE, we mention the  C++ libraries \texttt{HElib}~\cite{halevi2014algorithms}, \texttt{SEAL}~\cite{sealcrypto}, and \texttt{PALISADE}~\cite{palisade}. Due to the high complexity of the implemented schemes, these libraries require expert users. However, they provide a large variety of capabilities. One important capability, which can be decisive when judging the realizability of encrypted control, is the possibility of effectuating the encrypted computations on native 64 bit arithmetic of current processors, rather than expensive multiprecision arithmetic. We also refer to the following compilers for fully homomorphic cryptosystems~\cite{Dathathri2019chet,Boemer2019ngraph} that automate the choice of the best parameters for the runtime and the best parallelization techniques.
Various software tools are also available in the realm of secure multi-party computation.
They implement, e.g., garbled circuits, oblivious transfer schemes, or secret-sharing protocols and they further allow to transform high-level code into these low-level secure protocols. The paper~\cite{hastings2019sok} surveys these tools and comments on their usability and availability.

We finally note that special hardware can achieve much better performance for some encrypted computations than commodity hardware. In fact, by hardwiring specific computations in GPUs and reconfigurable hardware such as FPGAs, a significant speed-up can be obtained. This path, that is a research area on its own, is taken by a plethora of works such as~\cite{Doroz2014accelerating,poppelmann2015accelerating,San2016efficient,Cousins2016designing,Roy2019fpga,Riazi2019heax}. In the scope of this paper, a custom digital design for encrypted control on an FPGA that makes use of hardware-optimized modular multiplications is proposed in~\cite{Tran2020implementing}.
\end{minipage}
}\vspace{-4mm}
\end{figure}

While the primer above presents basic functionalities of cryptosystems allowing encrypted computations, there are plenty of techniques in the literature that improve the efficiency and usability of these tools. We survey relevant cryptographic software libraries and advantageous hardware in the sidebar ``\hyperlink{side:SoftwareHardware}{Software for encrypted computations and hardware acceleration}''.

We end this technical section with a visual analogy. Typical encryption schemes can be envisioned as safe boxes, where a plaintext is stored safely and can be retrieved only by knowing the combination code. On the other hand, a scheme that allows encrypted computations, such as the ones presented in this section, can be viewed as a safe glovebox: one can manipulate the plaintext stored inside through the gloves (through ciphertext manipulation), without the need of retrieving it outside of the box using the combination code.

\section{Encrypted linear control}

As briefly discussed in the introduction, the first encrypted controllers implement linear control laws of the form
\begin{equation}
\label{eq:linearControl}
\ub(k)= \Kb \xb(k),
\end{equation}
where $\Kb \in \R^{m \times n}$. Such linear controllers are typically applied to linear (discrete-time) systems
\begin{equation}
\label{eq:linearSystem}
\xb(k+1)=\Ab \xb(k)+\Bb \ub(k)
\end{equation}
with $\Ab \in \R^{n \times n}$ and $\Bb\in \R^{n \times m}$ but, for the moment, the system dynamics are not of interest.
The first provably secure implementation of~\eqref{eq:linearControl} was realized in~\cite{Kogiso2015} using the multiplicatively homomorphic ElGamal scheme~\cite{ElGamal1985}.
We briefly note that encrypted implementations of \eqref{eq:linearControl} were even proposed slightly earlier in~\cite{Fujita2015} (and also in~\cite{Kogiso2015})  using Rivest-Shamir-Adleman (RSA) encryption~\cite{Rivest1978_RSA}. However, these implementations build on the multiplicative homomorphism of the so-called ``text-book'' RSA that is considered to be insecure (see, e.g.,~\cite[p.~412]{Katz2014}). 

In the following, we summarize three existing realizations of encrypted linear control that build on different cryptographic techniques. For completeness, we note that not only static linear controllers as in~\eqref{eq:linearControl} have been considered in the framework of encrypted control. In fact, a significant number of works (e.g.,~\cite{Kogiso2015,Kim2016,Cheon2018need,Teranishi2019stability,Murguia2020}) focus on the encryption of linear dynamic controllers of the form
\begin{subequations}
\label{eq:dynamicController}
\begin{align}
\nonumber\\[-9mm]
    \xib(k+1)= \Abc \xib(k) + \Bbc \yb(k), \\
    \ub(k)=\Cbc \xib(k) + \Dbc \yb(k),
\end{align}
\end{subequations}
where $\xib(k)$ and $\yb(k):=\Cb \xb(k)+\Db \ub(k)$ refer to the controller state and the system's output, respectively. Nevertheless, encrypting \eqref{eq:dynamicController} is significantly harder than~\eqref{eq:linearControl} since iterating the controller state in an encrypted fashion is more demanding as briefly explained in Section~``\ref{sec:robustnessStability}'' further below. We refer the interested reader to the original works for more details. 

\subsection{First implementation using ElGamal encryption}

The encrypted implementation of~\eqref{eq:linearControl} in~\cite{Kogiso2015} is based on the following two observations.
First, the control action  can be computed based on the matrix
\begin{equation}
\label{eq:matrixPhi}
\Phib(k) := \begin{pmatrix}
\Kb_{11} \xb_1(k) & \dots & \Kb_{1n} \xb_n(k) \\
\vdots & \ddots & \vdots \\
\Kb_{m1} \xb_1(k) & \dots & \Kb_{mn} \xb_n(k)
\end{pmatrix} \in R^{m \times n}
\end{equation}
that solely contains products of the controller parameters $\Kb_{ij}$ and the states $\xb_j(k)$.
In fact, $\ub_i(k)$ as in~\eqref{eq:linearControl} is obviously equivalent to  
\begin{equation}
\label{eq:uBasedOnPhi}
\ub_i(k) = \sum_{j=1}^n \Phib_{ij}(k) 
\end{equation}

\begin{wrapfigure}{l}{0.61\textwidth}
\vspace{-1mm}
\fcolorbox{black}{colorLightBlue}{
  \begin{minipage}{0.57\textwidth}
\textsf{\textbf{\hypertarget{side:Quantization}{Forming integer messages from quantized data}.}}
Most of the cryptosystems currently used for encrypted control offer integer messages spaces of finite size. Hence, in order to encrypt system states and to securely evaluate control laws, states and controller parameters need to be mapped onto these message spaces. The first step of this mapping is usually an element-wise approximation of the real-valued states and parameters 
with fixed-point numbers from the set 
$\Q_{\beta,\gamma,\delta}:=\left\{-\beta^\gamma, -\beta^\gamma+\beta^{-\delta},\dots, \beta^\gamma- 2 \beta^{-\delta},\beta^\gamma- \beta^{-\delta} \right\}$,
where $\beta,\gamma,\delta \in \N$ with $\beta \geq 1$ can be understood as the basis, the magnitude, and the resolution of the set, respectively.
For example, for $\beta=10$ and $\gamma=\delta=1$, we obtain 
$\Q_{\beta,\gamma,\delta} = \{-10,-9.9,\dots,9.8,9.9\}$.
The actual quantization is realized with a user-defined mapping $g: \R \rightarrow \Q_{\beta,\gamma,\delta}$ (e.g., rounding off or up) that provides fixed-point approximations of the form $\hat{x}:=g(x)$. 
We do not further specify the surjective mapping $g$ 
apart from the restriction that 
\begin{equation}
\label{eq:gCondition}
|g(x)-x|\leq \beta^{-\delta}  \quad \text{for every} \quad x \in [-\beta^\gamma,\beta^\gamma].
\end{equation}
In other words, the quantization error should be limited by the resolution for real-valued data in the range of $\Q_{\beta,\gamma,\delta}$. 
It remains to address the mapping from $\Q_{\beta,\gamma,\delta}$ to a suitable subset of the set of integers.
In this context, the central observation is that
$\beta^\delta \,\Q_{\beta,\gamma,\delta}  \subset \Z$, i.e.,
scaling $\Q_{\beta,\gamma,\delta}$ with the factor $\beta^\delta$ results in a subset of $\Z$. Now, it is often convenient to provide positive integers in a set of the form $\{0,\dots,\varphi-1\}$, where $\varphi$ is a user-defined parameter.
This can be realized based on the function $f:\Z \rightarrow \{0,\dots,\varphi-1\}$ with
$f(z):=z \,\,\modu \,\,\varphi$.
In fact, the combination of the quantization via $g$, the scaling with $\beta^\delta$, and the mapping $f$ leads to $f(\beta^\delta g(x)) \in \{0,\dots,\varphi-1\}$. The resulting integers can then be encrypted and processed through homomorphic operations. In this context, it is important to note that additions and multiplications are structurally preserved by scaling and the mapping $f$ (see, e.g., \cite[Eq.~(10)]{SchulzeDarup2019_CDC_2Clouds}).  Hence,  multiplicatively homomorphic ElGamal and  additively homomorphic Paillier can indeed be utilized to implement~\eqref{eq:PhiCloudElGamal} and \eqref{eq:uCloudPaillier}, respectively. Moreover, the scaling and mapping can be (partially) inverted and we refer to \cite[Sect.~II.C]{SchulzeDarup2019_CDC_2Clouds} for details on a partial inverse of $f$.
It is easy to see that inverting $g$, i.e., the quantization, is generally not possible. Hence, we have to deal with the corresponding quantization errors. As apparent from~\eqref{eq:gCondition}, such errors can be reduced by increasing $\beta$, $\gamma$, and $\delta$. 
However, the cardinality of $\Q_{\beta,\gamma,\delta}$ is simultaneously increased. Thus, when implementing an encrypted controller, the user has to tune these parameters to ensure a faithful quantization at a reasonable computation cost. 
\end{minipage}
}\vspace{-4mm}
\end{wrapfigure}

\noindent for every $i \in \{1,\dots,m\}$. Second, an encrypted cloud-based computation of \eqref{eq:matrixPhi} can be implemented using ElGamal. To this end, the current state $\xb(k)$ is measured and encrypted at the sensor. In the following, we use the shorthand notation $\lsem\cdot\rsem$ to highlight encrypted values. We note that the notation is slightly informal since it combines the mapping of the data onto the message space of the cryptosystem and the subsequent encryption without giving details on the actual implementation. We briefly present a suitable procedure in the sidebar~``\hyperlink{side:Quantization}{Forming integer messages from quantized data}'' and refer the interested reader to the respective original works.

The resulting ciphertexts $\lsem\xb(k)\rsem$ are then sent to the cloud, where the $m n$ products  
\begin{equation}
\label{eq:PhiCloudElGamal}
\lsem \Phib_{ij}(k) \rsem=\lsem \Kb_{ij} \rsem\otimes \lsem\xb_j(k)\rsem
\end{equation}
are evaluated in an encrypted fashion using an multiplicative homomorphism of the form~\eqref{eq:multiplicativelyHomomorphic}. 
The encrypted control scheme is completed by transmitting the encrypted matrix $\lsem\Phib(k)\rsem$ to the actuator, where it is decrypted and where the control action is computed according to~\eqref{eq:uBasedOnPhi}.

In alignment with Figure~\ref{fig:encryptedControl}, the core of an encrypted controller is the ability to compute encrypted control actions based on encrypted system states. Clearly, the previously summarized scheme shows this characteristic. The scheme additionally offers an encryption of the controller parameters~$\Kb$ as apparent from the computation of $\lsem\Phib(k)\rsem$.
The encrypted computation of $\lsem\Phib(k)\rsem$ builds on the multiplicative homomorphism of ElGamal. However, since ElGamal is (originally) not additively homomorphic, summing up the various rows of $\Phib(k)$ to obtain $\ub(k)$ has to be carried out at the actuator.  As a consequence, the full matrix $\lsem\Phib(k)\rsem$, i.e., $m n$ ciphertexts, have to be transmitted from the cloud to the actuator. Taking into account that the original control scheme requires the transmission of only $m$ control inputs, the high communication load of the scheme becomes apparent. In addition, $mn$ decryptions are required at the actuator. Hence, this scheme also reveals more information to the actuator than the required $\ub(k)$. However, it is important to note that the control gain $\Kb$ remains private.

\subsection{Alternative implementation using Paillier encryption}

The high communication load and the high number of decryptions can be avoided by using the additively homomorphic Paillier cryptosystem instead of the multiplicative homomorphic ElGamal encryption. This observation was first made and implemented in~\cite{Farokhi2016}. 
As before, the proposed scheme calls for encryptions of the system states at the sensor and the resulting ciphertexts $\lsem\xb(k)\rsem$ are again sent to the cloud. However, by exploiting the homomorphism~\eqref{eq:additivelyHomomorphic}~or, more precisely, \eqref{eq:addPaillier} and~\eqref{eq:mulCPaillier},
 the scheme directly computes encrypted inputs in the cloud via
\begin{equation}
\label{eq:uCloudPaillier}
\lsem\ub_i(k)\rsem=\left(\Kb_{i1} \odot \lsem\xb_1(k)\rsem  \right) \oplus \dots \oplus \left(\Kb_{in} \odot \lsem\xb_n(k)\rsem  \right)
\end{equation}
instead of the intermediate result $\lsem\Phib(k)\rsem$ as above.  The encrypted inputs $\lsem\ub(k)\rsem$ are then sent to the actuator, where the decryption is carried out with the help of~\eqref{eq:decPaillier}.

The presented encrypted controllers based on ElGamal and Paillier encryption, respectively, offer many similarities. First, they both implement the same control law using HE. A second similarity becomes apparent when analyzing the intermediate results $\lsem\Psib_{ij}(k)\rsem:=\Kb_{ij} \odot \lsem\xb_j(k)\rsem$ in \eqref{eq:uCloudPaillier}. In fact, decrypting the corresponding ciphertext matrix $\lsem\Phib(k)\rsem$ would result in the same plaintext matrix as the decryption of $\lsem\Psib(k)\rsem$ from~\eqref{eq:PhiCloudElGamal}.
The two schemes differ, however, in terms of the pre- and post-processing of intermediate results. In contrast to $\lsem\Phib(k)\rsem$, $\lsem\Psib(k)\rsem$ is computed without encrypting $\Kb$ which, eventually, allows to determine $\lsem\ub(k)\rsem$ in the cloud. As a consequence, the communication load and the amount of required decryptions at the actuator are reduced. Hence, if encrypting $\Kb$ is not required, encrypted linear control can be more efficiently implemented using Paillier. Nevertheless, implementing encrypted control using any homomorphic cryptosystem is numerically demanding. This is apparent from Table~\ref{tab:compTimesElGamalPaillier} that lists computation times and operation counts for encryptions, encrypted operations, and decryptions using the ElGamal and Paillier cryptosystem, respectively. Ultimately, the comparison between the two implementations points out the important trade-off between privacy and efficiency: concealing more information, such as both the control gains and system states, incurs more communication cost than concealing only the states.

\vspace{-2mm}
\begin{table}[h]
\caption{Computation times (in milliseconds) for a single encryption, encrypted operation, or decryption using the ElGamal or Paillier cryptosystem, respectively, with different (public) key lengths. The listed times reflect average run times for Python procedures running on a 2.7 GHz Intel Core i7-7500U. The ElGamal encryption has been implemented according to~\cite[pp.~400-401]{Katz2014}. The listed periods of usage for the different keys reflect the recommendations~\cite{NIST2016} of the US National Institute of Standards and Technology (NIST). The last row summarizes the operation counts for the encrypted linear controllers for systems with $n$ states and $m$ control inputs. }
\label{tab:compTimesElGamalPaillier}
\begin{center}
\vspace{-2mm}
\begin{tabular}{ccccccccc}
\toprule
Key length & Usage & \multicolumn{3}{c}{ElGamal} & \multicolumn{4}{c}{Paillier} \\
\cmidrule(lr){3-5}\cmidrule(lr){6-9}
& (recommended) & $\Enc$ & $\mathtt{Mul}$ & $\Dec$ & $\Enc$ & $\mathtt{MulByC}$  & $\mathtt{Add}$ & $\Dec$ \\
\midrule
1024 & outdated & 1.065 & 0.023 & 0.963 & \hphantom{0}1.523 & 0.017 & 0.021 & 0.412\\
2048 & 2016--2030 & 6.191 & 0.024 & 5.828 & 10.511 & 0.055 & 0.064 & 2.887\\
3072 & beyond 2030 & 19.633 & 0.031 & 18.132 & 28.731 & 0.082 & 0.118 & 9.540 \\
\midrule
\multicolumn{2}{c}{Operation count} & $n$ & $m n$ & $m n$ & $n$ & $m n$ & $m (n-1)$ & $m$ \\
\bottomrule
\end{tabular}
\end{center}
\vspace{-3mm}
\end{table}

\subsection{Efficient implementation using secret sharing}

The computational effort can be significantly reduced if tailored controller architectures and cryptographic tools are considered.
As an illustrative example, we consider the encrypted linear controller in~\cite{SchulzeDarup2019_CDC_2Clouds} that is based on secret sharing and two-party computation. 
At the sensor, the current state $\xb(k)$ is measured, a random manipulation $ \rb(k)$ is generated, and the shares $\xb^{(1)}(k):=\xb(k)+\rb(k)$ and $\xb^{(2)}(k):=\rb(k)$ are specified. These shares are then sent to two non-colluding clouds (e.g., operated by two different cloud providers). There, the linear control law~\eqref{eq:linearControl} is evaluated for the respective share by computing
$$
\vb^{(1)}(k):= \Kb \xb^{(1)}(k)  \qquad \text{and}\qquad \vb^{(2)}(k):= \Kb \xb^{(2)}(k),
$$
respectively. The resulting values $\vb^{(1)}(k)$ and $\vb^{(2)}(k)$ are independently transmitted to the actuator, where the control action is determined as $\ub(k)=\vb^{(1)}(k)-\vb^{(2)}(k)$. Since the manipulation $\rb(k)$ is newly generated in each time step $k$, the individual shares $\xb^{(1)}(k)$ and $\xb^{(2)}(k)$ contain no information on the actual state $\xb(k)$ (if some additional technicalities are considered). Likewise, $\vb^{(1)}(k)$ and $\vb^{(2)}(k)$ contain no information on $\ub(k)$. 

\begin{figure}[htp]
\centering
\includegraphics[width=0.96\textwidth]{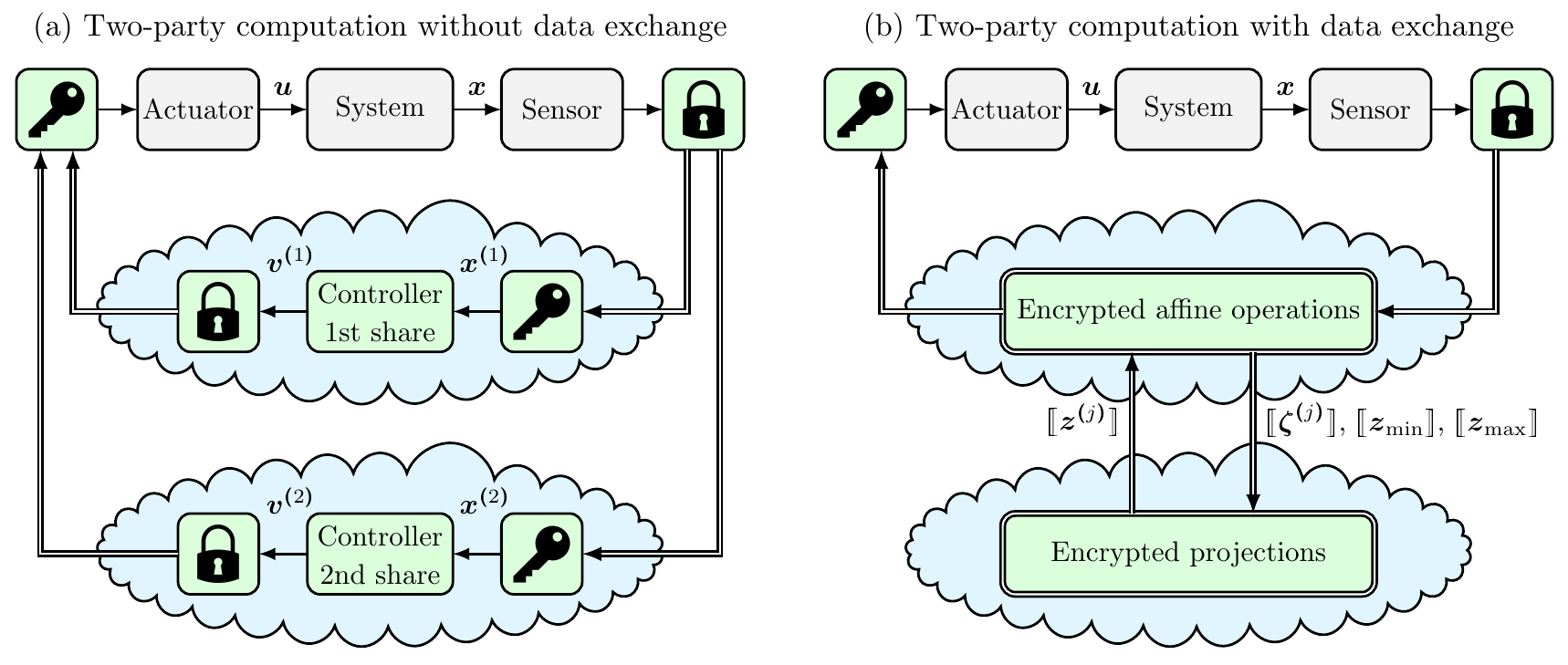} 
\caption{Illustration of (a) an encrypted control scheme based on two-party computation without data exchange between the clouds and (b) the encrypted implementation of a projected gradient scheme based on two non-colluding clouds with encrypted data exchange.}
\label{fig:twoCloudsSharing}
\end{figure}

The actual implementation in~\cite{SchulzeDarup2019_CDC_2Clouds} is based on one-time pads (that have been summarized above) and, similar to the implementations using homomorphic encryption, involves quantization of the states and controller parameters. 
It is important to understand the role of the two clouds.
In this context, we first note that involving two clouds that are not collaborating is realistic. Nevertheless, 
we need to guarantee that the first cloud has no access to the information transmitted to and from the second cloud and vice versa. Fortunately, this can be easily realized by establishing secure communication channels using standard symmetric encryption schemes (such as the advanced encryption standard (AES)~\cite{Daemen1999_AES}). The resulting scheme is illustrated in Figure~\ref{fig:twoCloudsSharing}.(a). Computation-wise, it is significantly more efficient than the previously discussed schemes since demanding asymmetric (i.e., public key) encryption and encrypted operations are not required. In fact, the needed symmetric encryption is relatively cheap. For instance, an AES encryption (and decryption) with a 128-bit key (that offers a security level comparable to the 3072-bit keys in Tab.~\ref{tab:compTimesElGamalPaillier}) requires less than a microsecond. As in the case of the solution involving the Paillier scheme, this solution considers public control gains. 

\section{Encrypted model predictive control}

Encrypted linear controllers serve as an intuitive proof of concept for encrypted control. The previously discussed schemes illustrate that encrypted cloud-based control can be realized ``naturally'' using existing cryptographic tools. However, cloud-based implementations of linear control laws are of limited use for real-world applications.
Indeed, using cloud services is more useful for numerically  demanding control strategies such as optimization-based control.
A popular control scheme that involves optimization is model predictive control (MPC). MPC for linear systems~\eqref{eq:linearSystem} typically builds on solving an optimal control problem (OCP) of the form 
\begin{align}
\label{eq:OCP}
V(\xb) := \min_{\substack{\tilde{\xb}(0),...,\tilde{\xb}(N)\\ \tilde{\ub}(0),...,\tilde{\ub}(N-1)}}   \tilde{\xb}(N)^\top \Pb \tilde{\xb}(N) &+ \sum_{\kappa=0}^{N-1} \tilde{\xb}(\kappa)^\top \Qb \tilde{\xb}(\kappa) +  \tilde{\ub}(\kappa)^\top \Rb  \tilde{\ub}(\kappa) \span \span \\
\nonumber
\text{s.t.} \qquad\qquad\qquad \tilde{\xb}(0) & = \xb,  \\
\nonumber
 \tilde{\xb}(\kappa+1)&=\Ab\,\tilde{\xb}(\kappa) + \Bb \tilde{\ub}(\kappa), && \forall \kappa \in \{0,\dots,N-1\}  \\
 \nonumber
 \tilde{\xb}(\kappa) &  \in \Xc, && \forall \kappa \in \{0,\dots,N-1\} \\
 \nonumber
 \tilde{\ub}(\kappa) & \in \Uc, && \forall \kappa \in \{0,\dots,N-1\} \\
 \nonumber
 \tilde{\xb}(N) & \in \Tc &&
\end{align}
in every time step for the current state $\xb = \xb(k)$. 
The OCP is characterized by the following parameters. The integer $N$ refers to the prediction horizon. The weighting matrices $\Pb\in \R^{n \times n}$, $\Qb\in \R^{n \times n}$, and $\Rb\in \R^{m \times m}$ specify the control objective. 
 The sets $\Xc \subseteq \R^n$ and $\Uc\subseteq \R^m$ describe state and input constraints of the system, respectively. Finally, the terminal set $\Tc \subseteq \Xc$ can be used to enforce stability of the closed-loop system and recursive feasibility of the MPC (see~\cite[Sect.~3.3]{Mayne2000} for details).
In this context, the closed-loop dynamics are determined
by the feedback 
\begin{equation}
\label{eq:uOptimal}
\ub(k)=\tilde{\ub}^\ast(0),
\end{equation}
where $\tilde{\ub}^\ast(0),\dots,\tilde{\ub}^\ast(N-1),\tilde{\xb}^\ast(0),\dots,\tilde{\xb}^\ast(N)$ refer to the optimizers for~\eqref{eq:OCP}.
For simplicity, we assume throughout the paper that $\Pb$, $\Qb$, and $\Rb$ are positive definite and that $\Xc$, $\Uc$, and $\Tc$ are convex  polyhedrons. Under these assumptions, \eqref{eq:OCP} is a strictly convex quadratic program (QP) with a unique optimizer.
Implementing MPC requires to solve this QP in every time step.
Depending on the size of $m$, $n$, and $N$ and the number of hyperplanes describing $\Xc$, $\Uc$, and $\Tc$, solving \eqref{eq:OCP} can be numerically demanding and a cloud-solution becomes beneficial.
Hence, encrypted cloud-based MPC offers interesting applications.

\subsection{First implementation using offline optimization}

The first encrypted MPC was introduced in~\cite{SchulzeDarup2018_LCSS}. The proposed implementation builds on the three following observations. First, the OCP \eqref{eq:OCP} can be rewritten as a QP of the form
\begin{align}
\label{eq:QP1}
\zb^\ast(\xb) = \arg \min_{\zb}\, \frac{1}{2} \,\zb^\top \Hb \zb &+ \xb^\top \Fb^\top  \zb \\
\nonumber
 \text{s.t.} \qquad\,\,\,\, \Gb \zb &\leq  \Eb \, \xb + \hb
\end{align}
with the decision variables
\begin{equation}
\label{eq:zQP1}
\zb := \begin{pmatrix}
\tilde{\ub}(0)\\
\vdots \\
\tilde{\ub}(N-1)
\end{pmatrix} \in \R^{Nm}
\end{equation}
and the parameters $\xb=\xb(k)$, where we refer to~\cite[Sect.~2.1]{Bemporad2002} for details on the computation of the vector $\hb$ and the matrices $\Eb$, $\Fb$, $\Gb$, and $\Hb$.
Second, the parametrized solution of~\eqref{eq:QP1} is known to offer a certain structure.  In fact, the optimizer $\zb^\ast(\xb)$ is a piecewise affine function of the parameter $\xb$. 
Since we have 
\begin{equation}
\label{eq:uCzRelation}
\ub(k)=\Cb \zb^\ast(\xb(k)) \qquad \text{with} \qquad
\Cb := \begin{pmatrix}
\Ib_m & \Ob
\end{pmatrix} \in \R^{m \times Nm}
\end{equation}
according to \eqref{eq:uOptimal}, the feedback becomes likewise a piecewise affine function of the form
\begin{equation}
\label{eq:uPiecewiseAffine}
\ub(k)= \left\{ \begin{array}{ll}
\Kb^{(1)} \xb(k) + \bb^{(1)} & \mathrm{if} \quad \xb(k) \in \Pc^{(1)},\\
\qquad \vdots & \qquad \vdots\\
\Kb^{(s)} \xb(k) + \bb^{(s)} & \mathrm{if} \quad \xb(k) \in \Pc^{(s)}. \\
\end{array}\right.
\end{equation}
The individual affine segments are specified by the matrices
$\Kb^{(\sigma)} \in \R^{m \times n}$, the vectors $\bb^{(\sigma)} \in \R^{m}$, and the polyhedra $\Pc^{(\sigma)} \subseteq \Xc$.
The number of segments~$s$ is limited by the (typically conservative) upper bound $2^{\vartheta}$ (see~\cite[Sect.~4.4]{Bemporad2002}), where $\vartheta$ refers to the number of inequalities in~\eqref{eq:zQP1}. 
Third, for moderate sizes of $m$, $N$, and $\vartheta$, the parameters  $\Kb^{(\sigma)}$, $\bb^{(\sigma)}$, and $\Pc^{(\sigma)}$ in~\eqref{eq:uPiecewiseAffine} can actually be computed offline, i.e., before runtime of the controller. A suitable procedure, which can be understood as a multi-parametric active-set solver, is described in~\cite[Sect.~4.3]{Bemporad2002}. Now, based on \eqref{eq:uPiecewiseAffine}, one can implement MPC without online optimization. Computing the control action $\ub(k)$ is then (typically) realized by a two-stage procedure. First, the polyhedron $\Pc^{(\sigma)}$ that contains the current state $\xb(k)$ is identified.  Second, the corresponding affine segment is evaluated.

\begin{figure}[tp]
\centering
\includegraphics[width=0.96\textwidth]{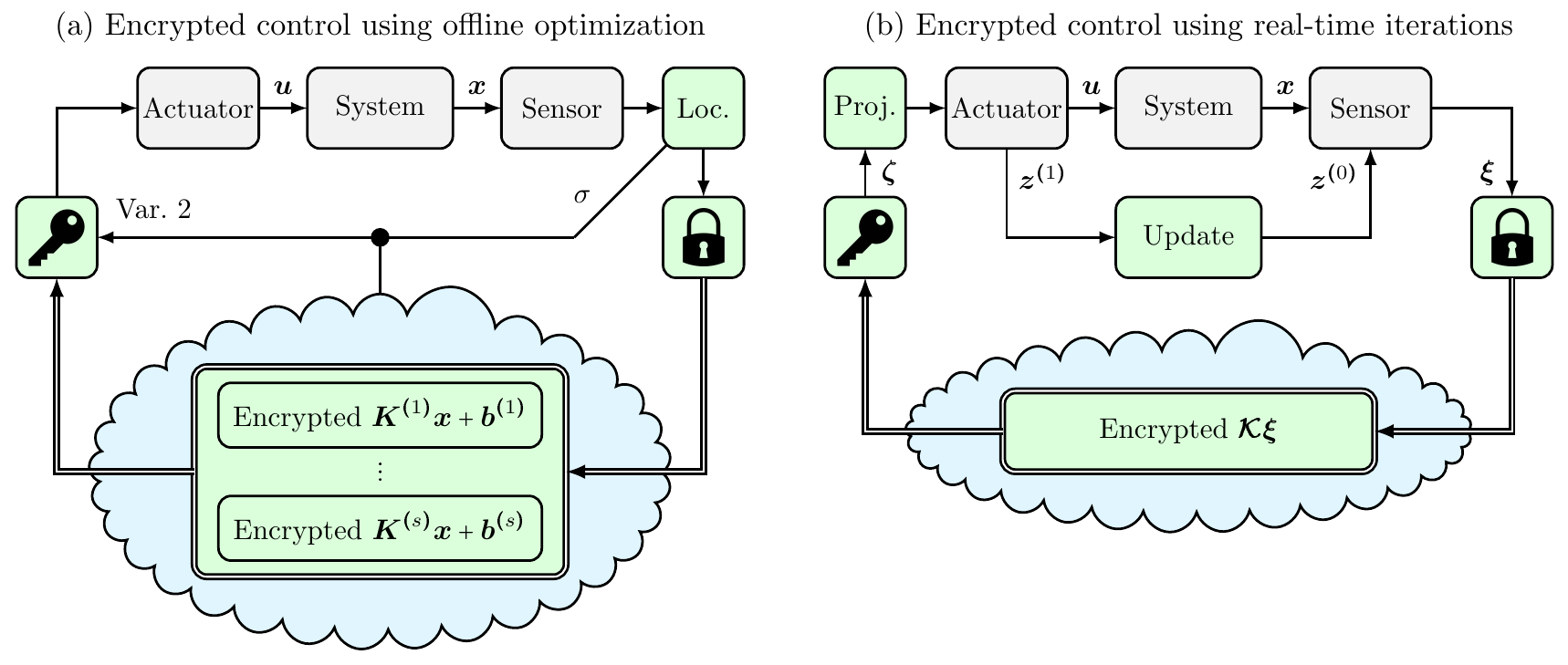} 
\caption{Illustration of encrypted MPC implementations based on (a) the parametric offline solution of the underlying optimal control problem and (b) real-time iterations of a projected gradient scheme.}
\label{fig:encryptedMPC}
\end{figure}

It easy to see that the second step of the procedure can be  implemented in an encrypted fashion. In fact, given the active segment in terms of the index $\sigma$, one can compute
\begin{equation}
\label{eq:uAffineCloud}
\lsem\ub(k)\rsem=\left(\Kb^{(\sigma)} \,\odot\, \lsem\xb(k)\rsem\right)  \oplus \lsem\bb^{(\sigma)}\rsem
\end{equation}
analogously to~\eqref{eq:uCloudPaillier} using Paillier. Here, with a slight abuse of notation, we extend the homomorphic operations ``$\odot$'' and ``$\oplus$'' to matrix-vector products and vector additions, respectively. We stress, in this context, that evaluating $\Kb^{(\sigma)} \,\odot\, \lsem\xb(k)\rsem$  also involves homomorphic additions (for $n>1$) as apparent from~\eqref{eq:uCloudPaillier}.
Hence, an encrypted MPC can be realized as follows. At the sensor, the current state $\xb(k)$ is measured and the corresponding $\Pc^{(\sigma)}$ is identified. Afterwards, $\xb(k)$ is encrypted and $\lsem\xb(k)\rsem $ is sent to the cloud together with the index $\sigma$. In the cloud, $\lsem\ub(k)\rsem $ is computed according to \eqref{eq:uAffineCloud} and transmitted to the actuator. There, $\lsem\ub(k)\rsem $ is decrypted and $\ub(k)$ applied. 
The resulting encrypted MPC is illustrated in Figure~\ref{fig:encryptedMPC}.(a).  We briefly note that robust MPC provides an effective way to deal with the quantization errors that appear during the preparation of~\eqref{eq:uPiecewiseAffine} for the encrypted implementation in~\eqref{eq:uAffineCloud} (see the sidebar~``\hyperlink{side:Quantization}{Forming integer messages from quantized data}'' for details). In fact, instead of the ``nominal'' OCP in~\eqref{eq:OCP}, one can consider the OCP underlying tube-based robust MPC and design the disturbance set such that it covers appearing quantization errors. 
Since the resulting OCP can again be rewritten as an QP of the form~\eqref{eq:QP1}, an encrypted implementation can be realized analogously to the scheme presented here (see \cite[Sect.~III]{SchulzeDarup2018_LCSS} for details).

We conclude the presentation of the scheme with a brief discussion of the information available to the cloud. Obviously, the clouds obtains the plaintexts $\Kb^{(1)}, \dots,\Kb^{(s)}$ and the ciphertexts $\lsem\bb^{(1)}\rsem,\dots,\lsem\bb^{(s)}\rsem$. During runtime, the cloud  additionally gets access to the current index $\sigma$ and accordingly selects the segment to be evaluated. In principle, knowing $\sigma$ could allow an intruder to narrow $\xb(k)$ down to $\Pc^{(\sigma)}$. However, the cloud has no direct access to the polyhedrons $\Pc^{(1)},\dots,\Pc^{(s)}$. 
Moreover, under the (realistic) assumption that the cloud has no information on the parameters of the OCP~\eqref{eq:OCP}, it seems hard to reconstruct $\Pc^{(\sigma)}$ from  knowledge of $\Kb^{(1)}, \dots,\Kb^{(s)}$. Nonetheless, if knowledge of $\sigma$ to the cloud is considered insecure, the sensor sends $\sigma$ to the actuator and not to the cloud (see Fig.~\ref{fig:encryptedMPC}.(a) for an illustration). Since the cloud has no information on the active segment, it has to evaluate \eqref{eq:uAffineCloud} for all $\sigma \in \{1,\dots,s\}$ and it has to transmit all resulting $\lsem\ub(k)\rsem$ to the actuator. There, only the ciphertexts corresponding to segment $\sigma$ are decrypted and applied. 
Comparing both variants, the latter offers more security. However, it is also less efficient since the computation effort in the cloud and the communication load to the actuator are significantly increased.
Apart from that, both variants share a common drawback. In fact, they both require $\Pc^{(\sigma)}$ to be identified at the sensor. There exist efficient algorithms for localizing $\Pc^{(\sigma)}$ that, for example, make use of binary search trees~\cite{Tondel2002}. Nevertheless, neglecting encrypted operations for the moment, identifying $\Pc^{(\sigma)}$ is significantly more demanding than evaluating the corresponding affine controller segment.
In summary, both variants of the proposed scheme show that encrypted cloud-based MPC can be realized. However, the scheme is only applicable for moderate $m$, $N$, and $\vartheta$. Moreover, the computational effort at the sensor is high. 

\subsection{Online optimization using real-time iterations}

The drawbacks of the previous scheme can be avoided by considering online optimization instead of evaluating~\eqref{eq:uPiecewiseAffine} offline. Several methods for solving QPs online exist. Examples include interior-point methods, active-set procedures, and proximal algorithms (see, e.g.,~\cite{Parikh2014}).
The various solvers have in common that they iteratively approach the optimal solution. We next summarize the encrypted MPC in~\cite{SchulzeDarup2018_NMPC} that builds on a projected gradient scheme (PGS), i.e., a structurally simple proximal algorithm. In order to apply the scheme, we introduce the set
$$
\Zc(\xb):=\left\{ \zb \in \R^{mN} \,|\, \Gb \zb \leq  \Eb \, \xb + \hb\,\right\}
$$
that reflects the state-dependent constraints in~\eqref{eq:QP1}. 
Then, a (primal) projected gradient descent can be realized by recurrently evaluating the iterations 
\begin{equation}
\label{eq:PGS_zIter}
\zb^{(j+1)} := \proj_\Zc \left( \zb^{(j)}- \rho \left(\Hb \zb^{(j)} + \Fb \xb \right)\right),
\end{equation}
where $\Hb \zb+ \Fb \xb$ is the gradient of the objective in~\eqref{eq:QP1}, where $\rho$ is the step size, and where $\proj_\Zc(\cdot)$ denotes a projection onto $\Zc(\xb)$. The choice of the parameter $\rho$ affects the convergence (or divergence) of the scheme. Convergence of the iterates $\zb^{(j)}$ to the optimizer $\zb^\ast(\xb)$ is guaranteed for every $\rho \in \left(0,2\lambda_{\max}^{-1}(\Hb)\right)$~\cite{Nesterov2013}, where $\lambda_{\max}(\Hb)$ refers to the largest eigenvalue of the positive definite Hessian $\Hb$.
Using the iterations~\eqref{eq:PGS_zIter} to solve~\eqref{eq:QP1} is only reasonable if the projection onto $\Zc(\xb)$ can be evaluated efficiently. This is, for example, the case if only box-shaped input constraints are present.
More precisely, this special case is characterized by 
\begin{equation}
\label{eq:onlyInputConstraints}
\Xc=\Tc=\R^n \qquad \text{and}\qquad \Uc=\{ \ub \in \R^m \,|\, \ub_{\min} \leq \ub \leq \ub_{\max} \},
\end{equation}
where $\ub_{\min} ,\ub_{\max}  \in \R^m$ reflect lower and upper bounds for the inputs, respectively. Under this assumption, we obtain
$\Zc(\xb)=\left\{ \zb \in \R^{mN} \,|\, \zb_{\min} \leq \zb \leq \zb_{\max}\,\right\}$ with
$$
\zb_{\min} := \begin{pmatrix}
\ub_{\min} \\
\vdots \\
\ub_{\min}
\end{pmatrix} \in \R^{Nm} \qquad \text{and} \qquad \zb_{\max} := \begin{pmatrix}
\ub_{\max} \\
\vdots \\
\ub_{\max}
\end{pmatrix} \in \R^{Nm}.
$$
The projection then yields~\eqref{eq:projectionBox}, which can, at least in plaintext, be efficiently evaluated.
\begin{equation}
\label{eq:projectionBox}
\proj_\Zc(\zetab)=\min \left\{ \max \{  \zb_{\min},\zetab \}, \zb_{\max} \right\}.
\end{equation}

Regarding the encrypted implementation, even the simplified projection remains demanding. In contrast, the computation of the projection's argument
$\zetab  := (\Ib_{Nm} - \rho \Hb) \zb^{(j)} - \rho  \Fb  \xb$ 
in~\eqref{eq:PGS_zIter} can be easily encrypted analogously to the Paillier-based linear scheme. This observation motivates the encrypted control scheme proposed in~\cite{SchulzeDarup2018_NMPC}. In this scheme, only a single solver iteration is evaluated per time step. More precisely, at step $k$, the cloud evaluates
\begin{equation}
\label{eq:encryptedRealtimeIter}
\lsem \zetab(k) \rsem = \left((\Ib_{Nm} - \rho \Hb) \,\odot\,\lsem \zb^{(0)}(k) \rsem \right) \oplus \left((- \rho  \Fb) \,\odot\,\lsem \xb(k) \rsem \right)
\end{equation}
based on encrypted states $\lsem \xb(k) \rsem$ and an encrypted initial guess $\lsem \zb^{(0)}(k) \rsem$ using the Paillier homomorphisms. The cloud then sends $\lsem \zetab(k) \rsem$ to the actuator, where it is decrypted and where the projection~\eqref{eq:projectionBox} is evaluated to obtain $\zb^{(1)}(k)=\proj_\Zc(\zetab(k))$.
Finally, the input $\ub(k)=\Cb \zb^{(1)}(k)$ is applied  analogously to~\eqref{eq:uCzRelation}. It remains to comment on the initial guess $\zb^{(0)}(k)$. The scheme in~\cite[Sect.~IV.A]{vanParys2018} suggests to reuse $\zb^{(1)}(k)$ to initialize $\zb^{(0)}(k+1)$. More precisely, $\zb^{(1)}(k)$ is forwarded from the actuator to the sensor, buffered, and then used to warmstart
\begin{equation}
\label{eq:warmstart}
\zb^{(0)}(k+1)=\Db \zb^{(1)}(k)
\end{equation}
 at the following sampling instance, where $\Db \in \R^{Nm \times Nm}$ is a user-defined update matrix. At step $k$, the sensor sends the encrypted initial guess $\lsem \zb^{(0)}(k) \rsem$ to the cloud along with  $\lsem \xb(k) \rsem$.
An illustration of the resulting encrypted controller is given in Figure~\ref{fig:encryptedMPC}.(b).

From a computational point of view, the previously described scheme is viable. However, from an optimal control perspective, the scheme seems questionable. 
Surprisingly, the coupling of solver iterations and sampling instances can perform well. The resulting real-time iterations work particularly well for MPC since the underlying OCPs for consecutive time steps are often very similar.
The good performance and certain stability certificates have even been proven analytically in~\cite[Sect.~IV]{vanParys2018}.
Clearly, the idea of encrypted real-time iterations can also be extended to solvers that can handle more complicated constraints than~\eqref{eq:onlyInputConstraints}, including state constraints. For instance, encrypted real-time iterations of the alternating direction method of multipliers (ADMM) are considered in \cite[Chap.~11.5]{SchulzeDarup2020}. 
Interestingly, the encrypted ADMM scheme involves encrypted updates of dual variables (or Lagrange multipliers) in the cloud, which causes the same issues as the encrypted realization of the dynamical controller in~\eqref{eq:dynamicController}. This relation is, however, not surprising since proximal algorithms such as PGS or ADMM can often be interpreted as dynamical controllers. 
 
\subsection{Online optimization using multi-party computation}

Focusing on the computations \eqref{eq:encryptedRealtimeIter} carried out in the cloud, the encrypted real-time iterations can be interpreted as  encrypted linear control for an augmented system state (consisting of $\xb$ and $\zb^{(0)}$). Hence, the utility of the cloud is again questionable. Moreover, real-time iterations may be stabilizing but they are often suboptimal. Both drawbacks can be compensated with the encrypted MPC proposed in~\cite{Alexandru2018_CDC}. In fact, \cite{Alexandru2018_CDC} presents a scheme based on two non-colluding clouds (illustrated in Fig.~\ref{fig:twoCloudsSharing}.(b)) that allows to evaluate multiple projected gradient iterations per time-step in an encrypted fashion. Thus, the computational load in the cloud(s) is increased, while the sensor and the actuator only handle the necessary encryption and decryption of data, respectively. In addition, multiple solver iterations generally increase the accuracy of the approximations. Convergence to the optimum $\zb^\ast(\xb)$ is further improved in~\cite{Alexandru2018_CDC} by implementing a fast gradient method.  However, for simplicity, we next present the central ingredients of the scheme for the PGS considered above.

Following the concept in~\cite{Alexandru2018_CDC}, the encryption of the PGS iterations~\eqref{eq:PGS_zIter} can be realized as follows. The first cloud (or server) is concerned with the encrypted evaluation of the affine argument of the projection. Hence, it computes
$$
\lsem\zetab^{(j+1)}(k)\rsem^{(2)} =  \left(\left (\Ib - \rho \Hb \right) \,\odot\, \lsem\zb^{(j)}(k) \rsem^{(2)} \right) \oplus \left((-\rho \Fb) \,\odot\, \lsem\xb(k)\rsem^{(2)}\right)
$$
analogously to~\eqref{eq:encryptedRealtimeIter}, where it makes use of an additively homomorphic cryptosystem that has been set up by the second cloud as indicated by the superscript ``$(2)$''. Hence, the second cloud provides the public key allowing encryption and homomorphic operations and it holds the secret key required for decryption.
Now, in order to complete an iteration $j$, one needs to perform the projection $\zb^{(j+1)}(k)=\proj_{\Zc}\left(\zetab^{(j+1)}(k)\right)$ in an encrypted manner. According to~\eqref{eq:projectionBox}, the projection onto the box-constraints $\Zc$ can be separated into the two steps
\begin{subequations}
\begin{align}
\label{eq:zetaMax}
\overline{\zetab}^{(j+1)}(k)&:=  \max \left\{  \zb_{\min},\zetab^{(j+1)}(k) \right\} \qquad \text{and} \\
\label{eq:zetaMin}
 \zb^{(j+1)}(k)&= \min \big\{   \overline{\zetab}^{(j+1)}(k), \zb_{\max} \big\}.
\end{align}
\end{subequations}
We explain the encrypted evaluation of~\eqref{eq:zetaMax} and omit the analogous handling of~\eqref{eq:zetaMin}. The procedure to be presented makes use of a protocol for private comparison~(see \cite{DGK07} for details) that we briefly summarize in the following. In this context, consider a party that holds two ciphertexts $\lsem z_0\rsem$ and $\lsem z_1\rsem$ encrypted with an additively homomorphic cryptosystem set up by a second party. As detailed in~\cite[Sect.~9.4.5]{Alexandru2020}, private comparison then allows the second party to learn a bit $\tau$ satisfying $z_\tau = \max \{z_0,z_1\}$ without revealing $\tau$ to the first party or the plaintexts $z_0$ and $z_1$ to any of the two parties. Using this protocol, the encrypted evaluation of~\eqref{eq:zetaMax} can be realized within five steps:

\begin{enumerate}
	\item The first cloud encrypts $\zb_{\min}$ using the public key from the second cloud (or receives $\lsem \zb_{\min} \rsem^{(2)}$ from the sensor) and randomly assigns the elements $\lsem \zetab_i^{(j+1)}(k) \rsem^{(2)}$ and $\lsem \zb_{\min,i}\rsem^{(2)}$ to $ \lsem \ab_i \rsem^{(2)}$ and $ \lsem \bb_i \rsem^{(2)}$ in order to mask their order.

\item The two clouds run a private comparison protocol on $ \lsem \ab \rsem^{(2)}$ and $ \lsem \bb \rsem^{(2)}$ that provides a vector $\taub \in \{0,1\}^{Nm}$ (such that $\taub_i=1$ whenever $\ab_i \leq \bb_i$)  to the second cloud.

\item The first cloud generates random vectors $\rb$ and $\ssb$ and masks $ \lsem \ab \rsem^{(2)}$ and $ \lsem \bb \rsem^{(2)}$ by computing $ \lsem \ab \rsem^{(2)} \oplus  \lsem \rb \rsem^{(2)}$ and $ \lsem \bb \rsem^{(2)} \oplus  \lsem \ssb \rsem^{(2)}$. The resulting ciphertexts $ \lsem \ab + \rb \rsem^{(2)}$ and $ \lsem \bb + \ssb \rsem^{(2)}$ are sent to the second cloud.

\item The second cloud defines $ \lsem \cb_i \rsem^{(2)}$ as $\lsem \ab_i + \rb_i \rsem^{(2)} \oplus \lsem 0 \rsem^{(2)}$ whenever $\taub_i=0$ and ${\lsem \bb_i + \ssb_i \rsem^{(2)}} \oplus \lsem 0\rsem^{(2)}$ otherwise, where the addition of $\lsem 0 \rsem^{(2)}$ implies a refreshment of the encryption. The ciphertexts $ \lsem \cb \rsem^{(2)}$ are then returned to the first cloud together with $ \lsem \taub \rsem^{(2)}$.

\item The first cloud computes for every $i \in \{1,\dots,Nm\}$
\begin{equation}
\label{eq:zetaBarEnc}
\big\lsem \overline{\zetab}_i^{(j+1)}(k) \big\rsem^{(2)} =   \lsem \cb_i \rsem^{(2)} \oplus \left(\rb_i \,\odot\, \left(\lsem \taub_i \rsem^{(2)}\right)  \oplus \lsem -1 \rsem^{(2)} \right) \oplus \left((-\ssb_i) \,\odot\, \lsem \taub_i \rsem^{(2)} \right).
\end{equation}
	
\end{enumerate}

The three last steps represent a variant of the oblivious transfer introduced in the ``Basics" section. In particular, \eqref{eq:zetaBarEnc} is a variant of \eqref{eq:obliviousTransfer}.
The computation of $\lsem \zb^{(j+1)}(k) \rsem^{(2)}$ based on $\lsem \overline{\zetab}^{(j+1)}(k) \rsem^{(2)}$ and $\lsem \zb_{\max}\rsem^{(2)}$ can be realized analogously. After evaluating $J \geq 1$ iterations, the first cloud holds $\lsem \zb^{(J)}(k) \rsem^{(2)}$, i.e., an encrypted approximation of $\zb^\ast(\xb(k))$. In order to provide $\ub(k)=\Cb\zb^{(J)}(k)$ to the actuator, the encryption has to be changed from $\lsem \cdot \rsem^{(2)}$ to $\lsem \cdot \rsem^{(\mathrm{act})}$, where the superscript ``$(\mathrm{act})$'' stands for a cryptosystem set up by the actuator (or the system operator). 
The standard way for the first cloud to obtain $\lsem \ub(k) \rsem^{(\mathrm{act})}$ from $\lsem \ub(k) \rsem^{(2)}$ is to send to the second cloud a masked version $\lsem \ub(k) + \tb\rsem^{(2)}$. The second cloud uses its secret key to decrypt, and then the public key of the actuator to encrypt $\lsem \ub(k) + \tb\rsem^{(\mathrm{act})}$. The first cloud then subtracts the mask to obtain the desired value: $\lsem \ub(k) \rsem^{(\mathrm{act})} = \lsem \ub(k) + \tb\rsem^{(\mathrm{act})} \oplus \lsem - \tb\rsem^{(\mathrm{act})}$.

It remains to comment on the performance of the scheme. First of all, due to the multiple solver iterations, the scheme usually performs better than the previously presented controller based on real-time iterations. However,  multiple iterations may also increase the effect of quantization errors as specified in \cite[Sect.~VI]{Alexandru2018_CDC}.
Still, robustness can be guaranteed for a suitable  design of the quantization. Regarding the computational effort of the scheme, we note that the communication load is substantial, in particular for the comparison protocol in step 2). In fact, let $l$ be the number of bits required to represent  $ \lsem \ab \rsem^{(2)}$ (or $ \lsem \bb \rsem^{(2)}$), then the necessary communication rounds between the two clouds are of order $l$ per solver iteration.

\section{Encrypted distributed control}

So far, the paper has focused on centralized cloud-based controllers. For many large-scale applications, centralized control strategies are not tractable. For such systems, distributed control schemes are often the method of choice.
Distributed control is particularly useful if the system of interest consists of numerous subsystems with local actuating and sensing elements. In the simplest case, the $M\in \N$ subsystems show linear dynamics of the form
$$
\xb^{(i)}(k+1)=\Ab^{(ii)} \xb^{(i)}(k) + \Bb^{(i)} \ub^{(i)}(k) + \sum_{j=1,i\neq j}^M \Ab^{(ij)} \xb^{(j)}(k),
$$
where $\Ab^{(ii)} \in \R^{n_i \times n_i}$ and $\Bb^{(i)} \in \R^{n_i \times m_i}$ express local dynamics and where the various $\Ab^{(ij)} \in \R^{n_i \times n_j}$ (for $i\neq j$) represent physical couplings between the subsystems. Importantly, distributed control usually aims for a common control goal, e.g., a constant power frequency in a smart grid. Realizing such a goal through local control actions $\ub^{(i)}$ requires cooperation and hence communication between the agents. In this context, an agent represents one subsystem and the corresponding local sensor, control, and actuator (see Fig.~\ref{fig:encryptedCooperation}). 
In general, communication will not take place between every pair of agents. In fact, the graph describing the communication network is usually incomplete. We assume that all communication links are bidirectional and that every pair of agents can, in principle, interchange information through a direct link or via other agents. In addition, we assume that no parallel links exist.
In other words, we consider a communication graph that is undirected, connected, and simple. More formally, the communication graph $\Gc(\Vc,\Ec)$ is characterized as follows. 
The vertex set $\Vc:=\{1,\dots,M\}$ and the edge set $\Ec \subset \Vc \times \Vc$ represent the $M$ agents and the communication links between them, respectively. If $(i,j) \in \Ec$ then the agents $i$ and $j$ (with $i\neq j$) are communicating with each other and called neighbors. 
The set $\Nc_i:=\{ j \in \Vc \,|\,(i,j) \in \Ec \}$ collects all neighbors of agent $i$. The bidirectionality of the communication then implies $i \in \Nc_j$ whenever $j \in \Nc_i$ (for all $(i,j) \in \Ec$).

\subsection{Cooperative control using sparse feedback}

Any cooperative control scheme has to comply with the communication network. 
In this context, a structurally simple approach is a sparse linear control scheme of the form
\begin{equation}
\label{eq:sparseLinearControl}
\begin{pmatrix}
\ub^{(1)}(k)\\
\vdots \\
\ub^{(M)}(k)
\end{pmatrix}= \begin{pmatrix}
\Kb^{(11)} & \dots & \Kb^{(1M)}\\
\vdots  & \ddots & \vdots \\
\Kb^{(M1)} & \dots & \Kb^{(MM)}
\end{pmatrix} \begin{pmatrix}
\xb^{(1)}(k)\\
\vdots \\
\xb^{(M)}(k)
\end{pmatrix},
\end{equation}
where the off-diagonal block matrices $\Kb^{(ij)} \in \R^{m_i \times n_j}$ are zero matrices whenever $j \notin \Nc_i$, i.e., whenever no communication link exists.
The local controllers can then be written as
\begin{equation}
\label{eq:uiCooperative}
\ub^{(i)}(k) = \Kb^{(ii)} \xb^{(i)}(k) + \sum_{j \in \Nc_i} \Kb^{(ij)} \xb^{(j)}(k).
\end{equation}
While structurally simple, designing such distributed controllers is not trivial and we refer to~\cite{Lin2011} for details on tailored optimization-based design procedures.
As requested, evaluating $\ub^{(i)}(k)$ according to~\eqref{eq:uiCooperative} requires only information on the local state $\xb^{(i)}(k)$ and neighboring states $\xb^{(j)}(k)$. Typically, information on neighboring agents is provided by directly communicating the states $\xb^{(j)}(k)$ between the agents. This can be a security threat, if the local state information is sensitive and if neighboring agents are curious about this data. 

\subsection{Encrypted cooperative control}

Encrypted control allows to overcome this threat and to ensure confidentiality of the local states. In the following, we briefly summarize the implementation proposed in~\cite{SchulzeDarup2019_LCSS}. We note that encrypted cooperative control was previously realized in~\cite[Sect.~4]{Farokhi2017} for a randomized control scheme. The first step of the procedure in~\cite{SchulzeDarup2019_LCSS} does not involve any encryption.  
In fact, it simply proposes to compute the ``control shares'' 
\begin{equation}
\label{eq:vijControlShares}
\vb^{(ij)}(k):=\Kb^{(ij)} \xb^{(j)}(k)
\end{equation}
at agent $j$ and to communicate $\vb^{(ij)}(k)$ instead of $\xb^{(j)}(k)$. At agent $i$, the control action is then computed via
\begin{equation}
\label{eq:uiCooperativeWithShares}
 \ub^{(i)}(k) = \Kb^{(ii)} \xb^{(i)}(k) + \sum_{j \in \Nc_i} \vb^{(ij)} (k).
\end{equation}
instead of~\eqref{eq:uiCooperative}. 
This small modification already hides the local states $\xb^{(j)}(k)$ from agent $i$. Indeed, without any information on $\Kb^{(ij)}$ and $\xb^{(j)}(k)$ (including the dimension $n_j$) it is hard to reconstruct  $\xb^{(j)}(k)$ from $\vb^{(ij)}(k)$. Moreover, even if information on $\Kb^{(ij)}$ is available to or can be reconstructed by agent $i$, retrieving  $\xb^{(j)}(k)$ leads to an observability problem that requires further information on $\Ab^{(jj)}$, $\Bb^{(j)}$, and $\ub^{(j)}(k)$ (see \cite[Sect.~VI]{Alexandru2019_CDC} for details). 
The downside of the proposed modification is that agent $j$ would get access to the matrix $\Kb^{(ij)}$ and the control shares  $\vb^{(ij)}(k)$. Both quantities could disclose information on the control strategy of the neighboring agent $i$, which could likewise be a security threat. This threat can, however, be eliminated using HE. To see this, we first note that the computation of the control shares~\eqref{eq:vijControlShares} is very similar to the evaluation of the linear controller~\eqref{eq:linearControl}. Hence, an encrypted computation of $\vb^{(ij)}(k)$ can be realized by evaluating
\begin{equation}
\label{eq:vijEncrypted}
\lsem \vb^{(ij)}(k) \rsem^{(i)}= \lsem \Kb^{(ij)} \rsem^{(i)} \,\odot\, \xb^{(j)}(k)
\end{equation}
using additively HE similarly to~\eqref{eq:uCloudPaillier}, where the superscript ``$(i)$'' indicates that the public key of agent $i$ is used for encryption. As a consequence, multiple cryptosystems are applied simultaneously and we specify their interaction shortly. For the moment, we just point out that $\lsem \Kb^{(ij)} \rsem^{(i)}$ will be provided to agent $j$ before runtime by some trustworthy entity (and not by agent $i$). 
Now, comparing \eqref{eq:uCloudPaillier} and \eqref{eq:vijEncrypted}, we note that the two implementations differ in terms of 
the partially encrypted multiplications of the controller parameters and the system states. More precisely, in~\eqref{eq:vijEncrypted}, the controller parameters are encrypted and the states are plaintexts whereas the roles are swapped in~\eqref{eq:uCloudPaillier}. Nevertheless, both variants can be realized using an homomorphism of the form~\eqref{eq:mulCPaillier}. The differences between \eqref{eq:uCloudPaillier} and \eqref{eq:vijEncrypted} reflect slightly different setups. In fact, \eqref{eq:uCloudPaillier} is evaluated in the cloud which should not obtain any information on the system states. In contrast, \eqref{eq:vijEncrypted} is computed at agent $j$ who should not get access to $\Kb^{(ij)}$ (and who necessarily measures $\xb^{(j)}$ in plaintext). 
However, both setups coincide in that neither the cloud nor the agent should get information on the computed control actions or shares, respectively. 

\begin{figure}[htp]
\centering
\includegraphics[width=0.96\textwidth]{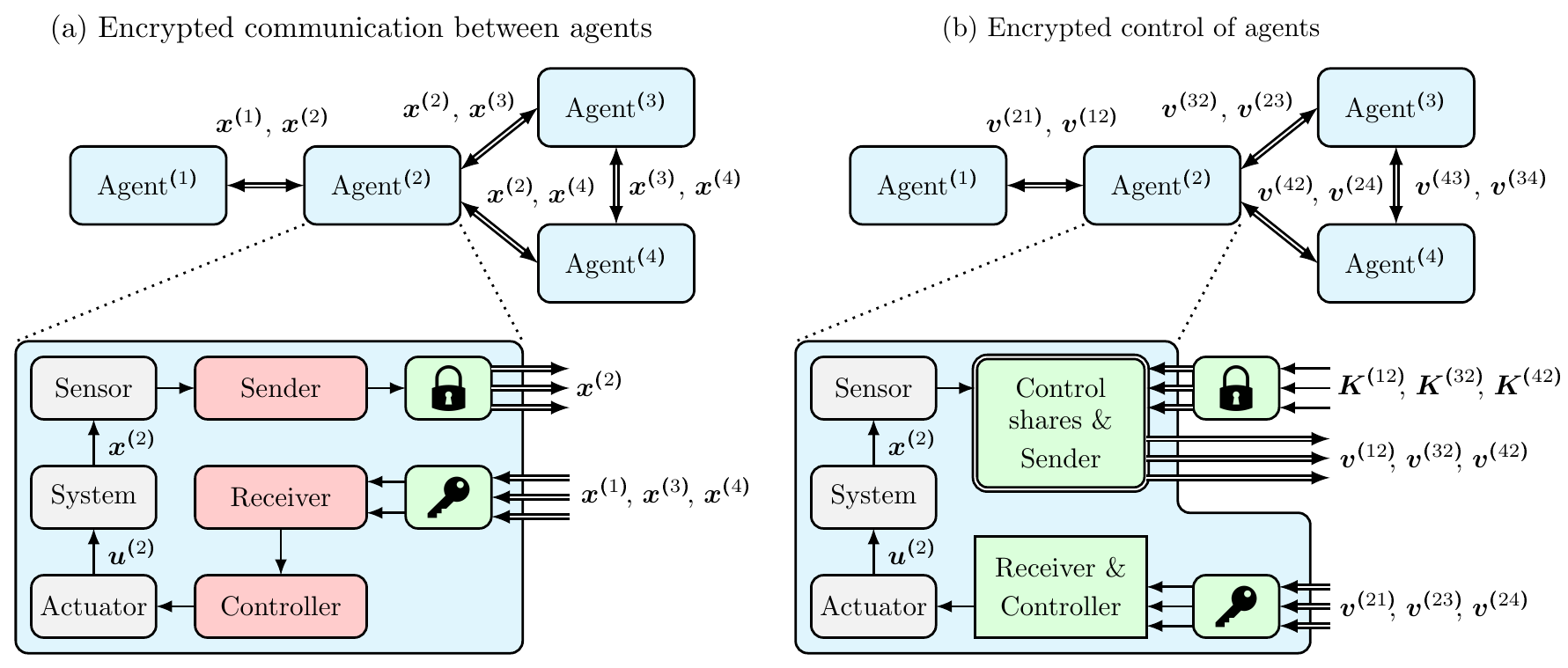} 
\caption{Illustration of (a) a cooperative multi-agent system with encrypted communications and (b) a multi-agent system operated by an encrypted cooperative control scheme. In (b),  control shares $\vb^{(ij)}=\Kb^{(ij)} \xb^{(j)}$ are computed, encrypted, and communicated instead of the classical exchange of neighboring states $\xb^{(j)}$ as in (a).}
\label{fig:encryptedCooperation}
\end{figure}

As briefly mentioned above, $\lsem \cdot \rsem^{(i)}$ stands for encryption with the public key of agent~$i$. As a consequence, only agent $i$ is capable of decrypting $\lsem \vb^{(ij)}(k) \rsem^{(i)}$ and, in principle, $\lsem \Kb^{(ij)} \rsem^{(i)}$, using its secret key. 
Since the multi-agent system consists of $M$ agents, $M$ cryptosystems are in use simultaneously. They are generated before runtime by the individual agents along with the corresponding secret keys. While the latter are kept private, the public keys are broadcast between the agents. Another item that has to be generated before runtime is the (sparse) control law~\eqref{eq:sparseLinearControl}. The corresponding procedure in~\cite{Lin2011} requires a centralized design. The resulting block matrices $\Kb^{(ij)}$ contain sensitive information. In fact, agent $j$ knowing $\Kb^{(ij)}$
 would render the encrypted computation~\eqref{eq:vijEncrypted} useless since computing $\vb^{(ij)}(k)$ in plaintext would then be trivial for agent~$j$.
 To overcome this issue, we assume that the centralized controller design is carried out by some trustworthy entity, e.g., a system operator. After computing the block matrices $\Kb^{(ij)}$, this entity repeats the following two steps for every agent $i \in \{1,\dots,M\}$. First, it sends $\Kb^{(ii)}$ to agent $i$ using a secure channel. Second, for every neighbor $j \in \Nc_i$, it encrypts $\Kb^{(ij)}$ using the public key of agent $i$ and sends the result $\lsem \Kb^{(ij)} \rsem^{(i)}$ to agent $j$.
During runtime, every agent runs data dissemination and assimilation phases in every time step $k$. During the data dissemination phase, every agent $j$ initially measures its own state $\xb^{(j)}$. It then computes the encrypted control shares $\lsem \vb^{(ij)}(k) \rsem^{(i)}$ according to \eqref{eq:vijEncrypted} for every $i \in \Nc_j$ and sends the result to the neighbor $i$. During the assimilation phase, every agent $i$ first receives $\lsem \vb^{(ij)}(k) \rsem^{(i)}$ from all the neighbors $j \in \Nc_i$. It then decrypts the shares using its secret key and computes the local control action according to~\eqref{eq:uiCooperativeWithShares}. 
The resulting encrypted cooperative control scheme is illustrated in Figure~\ref{fig:encryptedCooperation}.(b). 

\subsection{Improving security}

The key challenge in encrypted cooperative control is to ensure privacy of the local state $\xb^{(j)}(k)$ while being required to compute~\eqref{eq:uiCooperative}, i.e., the control action of the neighboring agent~$i$.
Providing $\vb^{(ij)}(k)$ instead of $\xb^{(j)}(k)$ to agent $i$ without revealing $\vb^{(ij)}(k)$ or $\Kb^{(ij)}$ to agent $j$ is a significant step towards a solution to this challenge. However, depending on the available information about the networked control system, it might be possible for agent~$i$ to reconstruct  $\xb^{(j)}(k)$ from  $\vb^{(ij)}(k)$ and data from previous time steps. 
In principle, there is an obvious way to render the reconstruction problem more difficult. In fact, providing the aggregated share
\begin{equation}
\label{eq:aggregatedShare}
\wb^{(i)}(k):=\sum_{j \in \Nc_i} \vb^{(ij)} (k)
\end{equation}
to agent $i$ instead of the individuals shares $\vb^{(ij)}(k)$ makes it significantly harder to reconstruct the individual $\xb^{(j)}(k)$ presupposed $|\Nc_i|\geq 2$. Consider, for example, the case of dynamically independent subsystems without physical couplings, i.e., $\Ab^{(ij)}=\Ob$ whenever $i\neq j$, and let $\Nc_i = \{j_1,\dots,j_{|\Nc_i|}\}$. Then, it is easy to show that the pair 
$$
\left(  \begin{pmatrix}
\Ab^{(i,j_1)} & & \Ob \\
& \ddots & \\
\Ob & & \Ab^{(i,j_{|\Nc_i|})}
\end{pmatrix}, \begin{pmatrix}
\Kb^{(i,j_1)} & \dots & \Kb^{(i,j_{|\Nc_i|})}
\end{pmatrix} \right)
$$
can be unobservable, which is desirable here, even if all individual pairs $(\Ab^{(ij)},\Kb^{ij)})$ are observable (see~\cite[Sect.~VII]{Alexandru2019_CDC} for details). 

Designing an efficient protocol that provides $\wb^{(i)}(k)$ to agent $i$ without revealing $\vb^{(ij)} (k)$ is not straightforward. A related problem, called private sum aggregation, has a long history of dedicated research~\cite{Shi2011privacy,Joye2013scalable,Benhamouda2016new,Bonawitz2017practical}. Conceptually, one strategy implemented in~\cite{Alexandru2019_CDC} is as follows. At every time step $k$, the neighbors of agent~$i$ agree upon some random shares $\rb^{(ij)}(k)$ that sum up to the zero vector in $\R^{m_i}$, i.e.,
\begin{equation}
\label{eq:randomShares}
\sum_{j \in \Nc_i} \rb^{(ij)} (k) = \Ob.
\end{equation}
More precisely, these shares should be in a finite field, as described in the section with secret sharing basics. 
All neighboring agents then send $\vb^{(ij)}(k)+\rb^{(ij)}(k)$ to agent $i$ instead of $\vb^{(ij)}(k)$. Now, although agent $i$ receives $|\Nc_i|$ individual shares, it is impossible to reconstruct the individual $\vb^{(ij)}(k)$ since the  random shares $\rb^{(ij)}(k)$ are unknown to agent $i$. However, by exploiting relation~\eqref{eq:randomShares}, summing up the received shares immediately leads to~\eqref{eq:aggregatedShare} and thus allows to evaluate~\eqref{eq:uiCooperativeWithShares}. 
Regarding the secure implementation, agent $j$ computes 
$$
\left(\lsem \Kb^{(ij)} \rsem^{(i)}\,\odot\, \xb^{(j)}(k) \right) \oplus \lsem \rb^{(ij)} (k) \rsem^{(i)}
$$
instead of \eqref{eq:vijEncrypted} and submits the resulting ciphertexts to agent $i$.
As a result, one obtains an encrypted control scheme that still hides neighboring control strategies in terms of $\Kb_{ij}$ using HE. In addition, masking by random values is used to keep the control shares $\vb^{(ij)}(k)$ of neighboring agents secret.

Not surprisingly, the improved security comes at a price. Preserving privacy requires different shares $\rb^{(ij)} (k)$ at every time step $k$. Recurrently generating these shares of zero is computationally demanding. One solution to this problem is to contract a trusted entity (e.g., the system operator) that generates the secret shares of zero for one (or more) time step(s) in advance and distributes them to the agents in the network. However, mainly depending on the size of the network, the random number generation can be very costly, both in terms of computation time and storage space. Alternatively, a decentralized method to generate the secret shares of zero can be enforced. As discussed in~\cite{Alexandru2019_CDC}, the neighboring agents of an agent $i$ then require two communication rounds to agree on suitable shares $\rb^{(ij)} (k)$ for an upcoming time step $k$.

\section{Quantization effects, stability guarantees, and continuous operation}
\label{sec:robustnessStability}

For simplicity of exposition, many technical details have been omitted during the presentation of the various encrypted control schemes above. 
In this section, we point out some important details for the correct functioning of encrypted control. 
First, the implementation of all presented encrypted control schemes requires to quantize, e.g., the system states $\xb_j(k)$ and controller parameters (such as $\Kb_{ij}$ for the linear schemes), in order to enable encryption or secret sharing. As discussed in the sidebar~``\hyperlink{side:Quantization}{Forming integer messages from quantized data}'', the quantization is irreversible and consequently leads to quantization errors. Carefully investigating such errors is essential since they may destabilize an originally stable control-loop (see, e.g.,~\cite{Delchamps1990,Istepanian2012digital}).
Hence, we illustrate quantization effects for linear controllers and note that an analogous analysis can be carried out for the other schemes.
With quantization involved, one actually applies 
\begin{equation}
\label{eq:uHatLinear}
\hat{\ub}(k):=\hat{\Kb} \hat{\xb}(k),
\end{equation}
instead of~\eqref{eq:linearControl}, where $\hat{\Kb}$ and $\hat{\xb}(k)$ refer to quantized parameters and states, respectively, and where  $\hat{\ub}(k)$ denotes the resulting control action. 
A classical approach is to interpret the quantization-based deviations $\hat{\ub}(k)-\ub(k)$ as disturbances (see, e.g., \cite[Sect.~III]{Delchamps1990}). 
Following this approach, that has been successfully applied to the encrypted control schemes in \cite{SchulzeDarup2018_LCSS,SchulzeDarup2018_NMPC,SchulzeDarup2019_LCSS,Alexandru2018_CDC}, the closed-loop dynamics for a linear plant~\eqref{eq:linearSystem} controlled by~\eqref{eq:uHatLinear} can be written as 
\begin{equation}
\label{eq:disturbedSystem}
\xb(k+1)=(\Ab+\Bb \Kb) \,\xb(k) + \Bb \db(k)
\end{equation}
with the state-dependent quantization errors 
\begin{equation}
\label{eq:disturbances}
\db(k):=\hat{\Kb} \hat{\xb}(k) - \Kb \xb(k) = \hat{\Kb} \left( \hat{\xb}(k) -  \xb(k) \right) + \left( \hat{\Kb} -  \Kb \right) \xb(k)
\end{equation}
that are regarded as disturbances hereafter.
The dynamics~\eqref{eq:disturbedSystem} can now be investigated using standard techniques from robust control. For example, under the assumption that the original control law~\eqref{eq:linearControl} has been designed such that $\Ab+\Bb \Kb$ is (Schur) stable, it is straightforward to certify input-to-state stability (ISS) with respect to $\db$ (see, e.g., \cite[Exmp.~3.4]{Jiang2001}). Moreover, additionally assuming bounded disturbances $\db(k)$, robust positively invariant (RPI) sets (see, e.g., \cite{Kolmanovsky1998}) can be considered.
Interestingly, we have
$$
\|\hat{\xb}(k) -  \xb(k)\|_\infty \leq \beta^{-\delta} \quad \text{and} \quad |\hat{\Kb}_{ij} -  \Kb_{ij} | \leq \beta^{-\delta}
$$
whenever $\|\xb(k)\|_\infty \leq \beta^\gamma$ and $|\Kb_{ij}| \leq \beta^\gamma$ according to \eqref{eq:gCondition} in the sidebar~``\hyperlink{side:Quantization}{Forming integer messages from quantized data}''. Hence, bounds for $\db(k)$ as in~\eqref{eq:disturbances} can be guaranteed and even kept arbitrarily small by suitably designing the quantization in terms of the parameters $\beta,\gamma,\delta \in \N$ and by restricting the attention to states in the set $\Xc_{\mathrm{res}}:=\{ \xb \in \R^n \,|\, \|\xb(k)\|_\infty \leq \beta^\gamma \}$.
In fact, the disturbances $\db(k)$ are then contained in the bounded set
$$
\Dc:=\left\{ \db = \hat{\Kb} \Delta \xb + \left( \hat{\Kb} -  \Kb \right) \xb \in \R^m \,\left|\, \|\Delta \xb \|_\infty \leq \beta^{-\delta},\, \xb \in \Xc_{\mathrm{res}} \right. \right\}.
$$
Now, a set $\Rc \subset \R^n$ is RPI for \eqref{eq:disturbedSystem} subject to the constraints $\xb(k) \in \Xc_{\mathrm{res}}$ and $\db(k) \in \Dc$ if
\begin{equation}
\label{eq:conditionRPI}
(\Ab + \Bb \Kb) \Rc + \Bb \Dc \subseteq \Rc  \quad \text{and} \quad \Rc \subseteq \Xc_{\mathrm{res}},
\end{equation}
where the set-valued addition refers to a Minkowski sum. Classical results on RPI sets (as, e.g., collected in~\cite{Kolmanovsky1998})
can now be used to investigate the closed-loop behavior of an encrypted control scheme. For instance, it is easy to see that every state trajectory of \eqref{eq:disturbedSystem} that starts in the maximal RPI set, i.e., the largest  set $\Rc$ satisfying~\eqref{eq:conditionRPI}, converges to the minimal RPI set \cite[Rem.~4.1]{Kolmanovsky1998} without leaving $\Xc_{\mathrm{res}}$. While the conservative interpretation of quantization errors as disturbances works sufficiently well for many encrypted control schemes, quantized control is an established field of research (see, e.g.,~\cite{Curry1970estimation,Nesic2009unified,Istepanian2012digital} for an overview) and more sophisticated approaches exist. Applying these techniques could be an interesting direction for future research. However, quantization effects are typically not the bottleneck for most existing encrypted controllers since quantization is a design-parameter. In other words, if the original control law is stabilizing then there usually exists a suitable quantization that  preserves (robust) stability. 

Depending on the nature of the cryptosystem underlying an encrypted controller, errors other than quantization errors can occur. For example, HE schemes based on the Learning with Errors hardness problem~\cite{Regev2010learning} require some noise to ensure confidentiality.
Similar to~the discussion above, such noise can be interpreted as a disturbance and its effects can~be~bounded accordingly. For a single encrypted operation, neither the quantization nor the noise  will significantly affect the computation.  
However, if the errors due to quantization or noise~accumulate, undesired effects may result. Such an accumulation appears, for example, for encrypted implementations of dynamical controllers as in~\eqref{eq:dynamicController}. In fact, iterating the encrypted controller states in the cloud usually implies an accumulation of quantization errors and noise, which might impede a continuous and seamless operation for an unbounded runtime. A continuous~operation can be easily realized by regular ``refreshments'' of the encrypted controller state through the client as, e.g., considered in \cite{Alexandru2019_CPS,SchulzeDarup2019_Chapter_ADMM,Murguia2020}. Since a ``refreshment'' usually implies~decryption and re-encryption, such solutions can sometimes be costly or inconvenient. 
More sophisticated schemes that ensure a continuous and seamless operation have been proposed in \cite{Kim2016,Cheon2018need,Kim2019encrypted,Teranishi2019stability}. In \cite{Kim2016}, the bootstrapping technique from fully HE is used to enable the cloud to carry out necessary refreshments on its own. However, since bootstrapping is still a very expensive operation,~\cite{Kim2016} employs three controllers operating at the same time for the same system.
In contrast, special transformations are considered in \cite{Cheon2018need,Kim2019encrypted} to ensure that the accumulated errors stay within certain limits. Finally, \cite{Teranishi2019stability} designed a stabilizing quantizer for a dynamic controller encrypted with the multiplicatively homomorphic ElGamal scheme, and~\cite{Murguia2020} discussed the reset of the dynamic controller's state using the additively homomorphic Paillier scheme.

\section{Current challenges and future research directions}

We presented various encrypted control schemes in order to illustrate different controller types, suitable architectures, and useful cryptographic tools. Although encrypted control is a young field of research, it is quickly emerging and the number of proposed schemes in the literature is already too large to be comprehensively covered in such an introductory article. Nevertheless, the presented selection illustrates two important characteristics of encrypted control. First, realizing encrypted controllers requires in-depth knowledge of both cryptography and control theory. In fact, encrypted control typically builds on suitable controller reformulations in combination with tailored cryptographic methods. Second, guaranteeing privacy usually increases both the computation effort and the communication load associated with networked control. This is, for example, apparent from the fact that homomorphically encrypted ciphertexts are typically significantly longer (measured in bits) than the corresponding plaintexts. For instance, cipthertexts resulting from the Paillier cryptosystem for a public key with 2048 bits have 4096 bits (cf. Tab.~\ref{tab:compTimesElGamalPaillier} and the sidebar ``\hyperlink{side:Paillier}{Paillier encryption and its homomorphisms}''). Regarding multi-party computation, the communication load is increased since the number of communicating parties typically growth compared to non-shared realizations.

The presented encrypted controllers also share some limitations that are common for existing schemes but that shall be overcome by future realizations.  First, partial homomorphic encryption (such as Paillier and ElGamal) and relatively simple multi-party computations are dominating the current generation of encrypted controllers and modern cryptosystems have not been frequently employed yet. 
Two exceptions to this observation are the controllers proposed in \cite{Kim2016} and \cite{Alexandru2019_CPS} that apply (only recently published) fully respectively leveled homomorphic encryption schemes. Second, existing encrypted controllers often 
rely on simplified or unrealistic setups. These simplifications include neglected latency, costly computations at sensors and actuators, and underused clouds. Third, most existing controllers require an ``unencrypted'' offline design of the scheme and privacy guarantees are exclusively given for the online computations.

In summary, the main bottleneck of encrypted control is the high additional computational effort and communication load. 
Hence, methods reducing this overload are of great interest. 
In this context, more realistic setups including, e.g., communication via lossy networks with latency, would further promote real-world applications like those in the sidebar ``\hypertarget{side:applications}{Prospective uses of encrypted control in industry}''. Moreover, from a control perspective, the encryption of more powerful controllers and robust stability certificates despite quantization errors and cryptograhic noise are needed.  
Finally, combining encrypted control schemes with other techniques ensuring \textit{security} (see, e.g., \cite{Teixeira2015_CSM}) and \textit{safety} (see, e.g., \cite{Ames2016control}) of control systems would be desirable.

The previous challenges indicate some interesting directions for future research on encrypted control. 
We conclude the paper with a list of concrete problems and we thereby invite researchers to contribute to this challenging and emerging field of research:

\begin{enumerate}
\item Propose encrypted implementations of more complex and nonlinear control schemes and realize the encrypted design of controllers or its parameters.

\item Reduce the computational effort and the communication load associated with encrypted control. Exploit, e.g., hardware acceleration and compilers for homomorphic encryption.

\item Consider quantization schemes that are not based on fixed-point numbers and further investigate the effect of quantizations errors. Provide robust stability certificates for encrypted control systems.  

\item Implement encrypted control in real-world applications such as smart grids or vehicular networks and verify its effectiveness. Investigate the effect of and counteract lossy communications and latency.

\item Consider ``active'' attacks (i.e., malicious adversaries) and couple encrypted control with complementary schemes focusing on integrity and availability of process data.

\item Ensure \textit{safety} of control systems without reducing \textit{security}. This is challenging since most known techniques ensuring safe control are highly nonlinear and require access to plenty of process data.

\item Use techniques from encrypted control to secure industrial applications such as Control-as-a-Service, smart grids, or transportation systems.

\end{enumerate}

Regarding the latter challenge, we briefly note that some encrypted computations are already deployed in industry.
For instance, distributed systems such as blockchains make use of cryptographic tools for coordination.
In particular, privacy-preserving frameworks such as Zcash~\cite{hopwood2016zcash} utilize heavy cryptography for performing transactions without leaking information. Furthermore, secure federated learning by Google~\cite{federatedGoogle,Bonawitz2017practical} exploits private aggregation to train a model from a multitude of different agents while reducing the amount of information collected. Such implementations focus on scalability, speed, and coordination, and could thus serve as guidance for deploying encrypted control in practice.

\section*{Acknowledgements}
M. Schulze Darup gratefully acknowledges the support by the German Research Foundation (DFG) under the grant SCHU 2940/4-1.

\end{document}